**Amateur astronomers and the new golden age of cataclysmic variable star astronomy**

**Jeremy Shears**

**Summary**

The study of cataclysmic variable stars has long been a fruitful area of co-operation between amateur and professional astronomers. In this Presidential Address, I shall take stock of our current understanding of these fascinating binary systems, highlighting where amateurs can still contribute to pushing back the frontiers of knowledge. I shall also consider the sky surveys that are already coming on stream, which will provide near continuous and exquisitely precise photometry of these systems. I show that whilst these surveys might be perceived as a threat to amateur observations, they will actually provide new opportunities, although the amateur community shall need to adapt and focus its efforts. I will identify areas where amateurs equipped for either visual observing or CCD photometry can make scientifically useful observations.

**Introduction**

It is often said that astronomy is one of the few remaining sciences where amateurs can still contribute to research and the study of variable stars is one of the fields regularly cited to exemplify that this is the case. In my 2016 Presidential Address, I reviewed how members of the BAA Variable Star Section (VSS), often equipped with only modest equipment, have contributed important observations that have helped push back the frontiers of variable star science for more than 126 years. Actually, the VSS observations database extends back to 1840, some 50 years before the Association was founded, and now contains nearly three million observations (1).

In this my second Presidential Address, I shall focus on one particular type of variable star, the Cataclysmic Variables (CVs). The study of CVs by amateur astronomers has underpinned breakthroughs in understanding the behaviours of these systems. Amateurs have been involved in the discovery and characterisation of new CVs and the VSS database contains some exquisite long-term visual light curves, which in recent years, as new technologies have emerged, have been supplemented by CCD photometry. Nowadays, it is common for amateurs and professionals to cooperate in this research and jointly to publish the results in the scientific literature. Whilst amateurs might have relatively small telescopes, they do have access to them whenever they chose (weather permitting!) and because they are located around the world at different longitudes, they can obtain near-continuous photometry. By contrast, professionals tend to have limited time on much larger telescopes, which is usually scheduled well in advance, but they are able to obtain more detailed astrophysical data and this is often supplemented with multi-wavelength observations from satellites. Therefore, the activities of professionals and amateurs are in fact complementary: working together across our community, both in obtaining the data and analysing it, has been common practise for many years. There exists a mutual trust and desire for understanding these systems that I believe is the key to the success of the collaboration.

However, new large surveys, such as Gaia, PanSTARRS and LSST, are coming online which will provide near constant monitoring of the skies, producing exquisitely precise photometry of millions of objects, including CVs. Does this sound the death-knell for amateur contributions to CV astronomy as our work is taken over by the all-seeing, untiring, automatic survey machines? "Not at all" is my response! Instead of putting us out of business, they will open up new opportunities and provide a further impetus for professional-amateur collaboration in what is emerging as a new golden age of variable star astronomy. On the other hand, we shall probably have to adapt and focus our efforts somewhat. So let us look at what CVs actually are and how amateurs contribute to their understanding.

## What are cataclysmic variables?

In astrophysical terms, CVs are compact binary stars comprising a white dwarf primary and a secondary that is usually a late-type main sequence star (Figure 1a). Because of the proximity of the two stars and the strong gravitational field of the white dwarf, the white dwarf distorts the shape of the secondary and draws material away from it. In the absence of a magnetic field, the material, which has substantial angular momentum, doesn't fall directly onto the primary, but instead forms an accretion disc through which it slowly spirals inwards towards the white dwarf. If on the other hand, the white dwarf has a strong magnetic field, an accretion disc cannot form. In this case, the accretion stream, which is mainly ionised hydrogen, follows the magnetic field lines and rains down onto the surface of the white dwarf near its magnetic poles (Figure 1b). There are also intermediate situations in which the magnetic field is not quite so strong, where a truncated accretion disc forms (Figure 1c).

CVs show variability on a wide variety of timescales (Table 1), but the most exciting events are outbursts. In the case of *novae*, the star can brighten by more than 10 million-fold in a matter of hours or days. The cause of this "cataclysm" is that material flowing through the accretion disk accumulates on the surface of the white dwarf and eventually causes a runaway thermonuclear reaction. The ensuing explosion causes the system to increase in brightness dramatically, blowing the outer layers of the white dwarf away into space as an expanding gas shell. With time, the gas cools and the once bright star begins to fade: the outburst is over. By contrast, the *dwarf novae*, as their name suggests, are less flamboyant. They brighten only about 100-fold and the origin of this more modest cataclysm lies within the accretion disc itself. As material builds up in the disc, a thermal instability is triggered that drives the disc into a hotter, brighter state. For a comprehensive review of CV's, the reader is directed to two excellent and highly readable texts on the subject by Brian Warner (2) and Coel Hellier (3).

## The significance of CVs as a physics laboratory

Accretion is a fundamental process and much research on CVs during the last half century has been on understanding the physics of accretion. Accretion discs are found in a wide variety of systems. For example, accretion occurs during star formation. In this process, as a star condenses out of a rotating interstellar cloud, the cloud gets flattened into a disc, which begins to be accreted onto the new star, forming a T Tauri object (Figure 2). When most of the material has been accreted, the remnants still carry significant angular momentum. These eventually condense into planets as the particles in the disc collide and stick together. This is probably how our solar system formed. On a larger scale, galaxies are initially formed as gaseous discs and many go on to develop central discs that fuel their active nuclei. In fact, as already mentioned, accretion discs form around black holes and it's thought there is an accreting supermassive black hole at the centre of many galaxies including our own.

Studying the accretion discs around black holes and in early stellar systems clearly poses severe observational challenges (and for our early Solar system it is impossible!), whereas CVs, because of their short timescales, provide a useful laboratory to study accretion disc physics. In their many guises, CVs provide an opportunity for us to probe a variety of discs and in the case of dwarf novae, where the disc dominates the light output, we can study those discs as they change between quiescence and outburst.

Some types of CV are also understood to be precursors of another astrophysically important object: the Type 1a supernovae, which are used as standard candles in determining the distance to the galaxies. The Type 1a supernova explosion occurs when the white dwarf grows to near the Chandrasekhar limit of approximately 1.4 times the mass of the sun. More about Type 1a supernovae later.

**A brief history of CVs** (4)

Novae have been noted throughout recorded history when a new star becomes visible in the night sky. Our ancient ancestors must have watched with wonder, and perhaps fear, when a new object suddenly appeared. However, dwarf novae entered onto the stage more recently, the first being U Geminorum which was discovered on 1855 December 15 by John Russell Hind (1823-1895; Figure 3a). Employed at George Bishop's observatory in Regent's Park (Figure 3b), Hind was primarily searching for minor planets at the time he stumbled across a $9^{th}$ magnitude object "shining with a very blue planetary light". The star soon faded below Hind's detection limit, but three months later it was again observed in outburst by Norman Pogson (1835-1891). Many outbursts of U Gem have been observed subsequently; the interval between outbursts varies widely with extremes of 62 days and 257 days.

U Gem and SS Cyg, another dwarf nova which was discovered in 1896 by Louisa D. Wells of Harvard College Observatory, were added to the VSS programme in 1904.SS Cyg spends most of its time in quiescence at $12^{th}$ magnitude, but every couple of months suddenly brightens to 8th magnitude for a few days before gradually fading again. One person who took a particular interest in SS Cyg in the early days of the VSS was Charles Lewis Brook (1855−1939; Figure 4) who was its third Director, serving from 1910 to 1921 (5). In 1911 Brook wrote: 'SS Cygni is a fascinating and mysterious star. We cannot hope to explain its changes without a complete record of its history. Towards this end the VSS may claim to have done its share during the past 5 years'. (6)

In spite of the lofty aims of the monitoring programme, by 1914 the true difficulty of the task of predicting the star's behaviour was becoming apparent to Brook. He commented: (7) 'It is to be feared that the available material is still insufficient and that we may have to wait some time to predict the star's movements'.

Brook was keen that variable star observers around the world should pool their observations to allow a more complete analysis of the star and he reached out to the AAVSO in a spirit of cooperation. By 1926, thoughts of predicting SS Cyg's outburst behaviour had all but faded, as Brook commented: 'I am aware that some attempts have been made to predict, but, as far as I am aware, none have succeeded.' (8)

We now know that outbursts of SS Cyg are, like all dwarf novae, only *quasi*-periodic, which means they cannot be predicted with certainty. Nevertheless, the observations submitted to the BAA VSS, as well as to other variable star organisations around the world, means that SS Cyg is one of the most intensively monitored of variable stars, with more than half a million observations. As a result, it appears that no outbursts have been missed since it was discovered more than 120 years ago. It shows repeated outbursts with a mean recurrence time of 49±15 days (9). A recent example which illustrates the importance of amateur surveillance of SS Cyg involves the measurement of the distance to SS Cyg – uncertainty in knowing this was troubling astronomers modelling its outbursts for decades. Definitive distance measurements were planned using very long baseline interferometric radio observations. Now, the radio outburst fades faster than the optical, so to catch this short-lived bright radio state and make high precision measurements, radio observations have to start as soon as possible after the outburst is detected. It therefore fell to amateur astronomers to report optical outbursts and these triggered radio observations over 10 epochs between 2010 April and 2012 October, finally allowing the definitive distance (114±2 pc) to be determined (10). Without the constant monitoring of amateur observers, the radio observations would not have been successful. Even today, amateurs continue to monitor the 'ups and downs' of SS Cygni and it is one of the most popular stars with new variable star observers because it is always doing something.

Perhaps the first golden age of CVs began in the 1960s (11) when it was confirmed that novae and dwarf novae are not single stars, but instead are binary systems. Evidence was gleaned largely from two observational techniques: spectroscopy and photoelectric photometry. Spectroscopic measurements were used to determine the radial velocities of the systems and sensitive photoelectric photometry, driven by the availability of the 1P21 photomultiplier tube from the mid-1940s, allowed real-time photometry with a resolution of less than a minute to be performed. This revealed characteristic features in the light curve, such as flickering and the existence of a "bright spot". Both features are now known to be due to the accretion stream: flickering is due to the stochastic nature of the accretion flow and the bright (or hot) spot occurs when the flow impacts the accretion disc (Figure 5). Photometry also showed that some of the systems were eclipsing binaries. Combining both techniques allowed new insights, as well as measurements of the orbital periods of the systems. For example, as part of a photometric survey of CVs, Merle Walker showed (12) that the remnant of Nova Herculis 1934 (discovered by BAA stalwart Manning Prentice on 1934 December 12), now called DQ Her, is an eclipsing binary with orbital period of 4.65 hours, similar to the nova-like variable UX UMa. By 1967, 23 CVs had known orbital periods, which ranged between 82 minutes and 227 days; 19 were spectroscopic binaries and 13 were eclipsing (13).

Further investigations in the 1960's by Kraft, Krzeminski, Mumford and others confirmed what we now know are the essential elements of a CV: the presence of a white dwarf which is accompanied by a late-type star filling its Roche lobe and an accretion disc. As more CVs have been discovered, amateur astronomers have played an important role in determining the long-term light curves of individual objects. The advent of low-cost CCD cameras in the 1990s has also allowed amateurs to engage in time-resolved photometry which has shed light of the properties of systems, in particular on characterising dwarf novae during their outbursts.

More recently, observations from orbiting X-ray satellites have shown that other binary systems exist which have some similar properties to CVs, such as an accretion discs, but which do not contain a white dwarf. Instead, the accreting object is a neutron star or a black hole. I shall have more to say about these later since amateur observations of these systems are also of great importance.

Whilst much is known about the astrophysics of CV's, this is still a very active area of research. Some of the outstanding questions that CV researchers are addressing are summarised in a paper by Paul Szkody and Boris Gänsicke (14).

**New challenges, new opportunities: amateur observations of CVs in the era of synoptic surveys**

Amateurs have contributed to CV research for well over a century. For much of this time, the observations have been visual and organisations like the BAA VSS and the AAVSO have accumulated millions of observations. Their observations are frequently used to study the long-term behaviours of CVs as well as to trigger follow-up observations with large telescopes or satellites. It was the century-long light curves of U Gem and SS Cyg compiled from amateur observations that catalysed the modelling of dwarf nova outbursts (15) and the simultaneous monitoring of SS Cyg by amateurs and orbiting X-ray satellites led by Peter Wheatley of Leicester University revealed an anti-correlation between hard and soft X-rays during an outburst (16), to cite only a couple of examples.

In the last two years there have been more than a dozen requests from professionals for assistance in monitoring stars in connexion with planned satellite observations. A recent example of amateurs supporting satellite observations took place in late 2016 and early 2017 when Dr. Christian Knigge (University of Southampton) requested coverage of the dwarf nova YZ Cnc. The aim was to trigger Target of Opportunity observations by the Chandra X-ray observations during a suitable outburst. As a result of an intensive monitoring campaign Chandra was successfully triggered during a outburst in mid-February 2017 (Figure 6).

Amateur CCD observations of some of the fainter dwarf novae have revealed that a few which were once thought to undergo rare outbursts actually outburst much more frequently. For example, CG Draconis was for many years believed to be an infrequently outbursting object until a campaign by VSS members revealed it actually "goes off" every 11 days on average! (17) Another campaign, this time on V1316 Cyg, revealed this CV has faint (~1.4 magnitude) outbursts every 10 days and these are of very short duration, lasting only a day or two at most. (18) There is some evidence that these are not normal dwarf nova outbursts, but stunted or attenuated outbursts involving excitation of only part of the accretion disc as their amplitude is only ~1.5 magnitudes. Because of the modest amplitude and short duration of these enigmatic events, they were only revealed by intensive monitoring by amateurs.

Being able to probe fainter detection limits has also been important in keeping pace with the discovery of fainter objects by sky surveys such as the Hamburg Quasar Survey, the All Sky Automated Survey (ASAS), the All-Sky Automated Survey for Supernovae (ASAS-SN), the Sloan Digital Sky Survey (SDSS) and the Optical Gravitational Lensing Experiment (OGLE). None of these was set up with the primary aim of discovering CVs, but their harvest of CVs has been bountiful. Amateurs have played an important role in characterising the CVs and countless papers have resulted; CVs identified by SDSS in particular have certainly kept the present author's clear nights occupied for more than a decade!

The new sky surveys that are planned will generate vastly more data and potentially interesting objects to observe that the professional community can hope to follow up. As Professor Gerry Gilmour told us in his 2015 BAA Christmas lecture, ESA's Gaia satellite (Figure 7a) was launched in 2013 with the aim of measuring accurate positions of a billion stars during its 5-year mission. Gaia is already discovering a whole host of CVs; new objects are announced on a website as well as via a smartphone "app" (Figure 7b) – on some days, the author's phone buzzes multiple times as new Gaia alerts are broadcast. And here is the rub: the Gaia team is actively soliciting amateur astronomers to provide follow-up time series CCD photometry. In fact, BAA member Dr Elmé Breedt of Cambridge University is part of the team which is enlisting the support of amateur astronomers. The reason being that Gaia does not provide sufficient time resolution to characterise these objects properly. Already this follow-up work is revealing new insights into CVs (an example is Gaia 14aae which is discussed later in the section on helium dwarf novae).

The planned earth-based synoptic surveys such as PanSTARRS and LSST will similarly provide many new targets for amateurs (Figure 8). LSST, for example, is due to come on-stream at the end of 2019 and even in the early stages is expected to discover about 100,000 variables per night and many of these will be CVs (19). Whilst they will likely have more intense sky coverage than earlier surveys, their cadence is not really suitable for CVs with orbital periods of hours and outbursts than can rise within one night. These seemingly all-seeing surveys have other disadvantages, for example they saturate at fairly bright limits, ~magnitude 16-18, which means many of the brighter CVs will still need to be monitored by other means, including visual. Moreover, the survey filters are generally at the red end of the spectrum, whilst CVs tend to be more active at the blue end.

There is also a question mark over how long each survey will remain operational, especially as funding pressures continue. The Gaia mission was planned for five years, although there is every hope that it will be extended by a further five. Time will tell about the longevity of other surveys, noting that some have yet to proceed further than the planning stage.

With these surveys in mind, let's look at the amateur's future contribution to CV science. Where can the amateur continue to play a role, what strengths should she play to and what opportunities are emerging? In the following sections we will consider these questions and provide examples of where amateurs are active at the cutting edge of CV research.

At this point it's worth reminding readers that the component stars, accretion disc, bright spot and geometry can all lead to variability - this is what makes CVs interesting systems to observe, if sometimes difficult to understand

**Long term monitoring programmes: VW Hyi**

The southern star VW Hyi was discovered photographically as a variable star by W.J. Luyten in 1932 (20) and was soon recognised as a dwarf nova, the second brightest at maximum in the class.  Visual observations of VW Hyi began in earnest in 1953, promoted by longstanding BAA member Albert Jones (1920-2013) under the auspices of the Royal Astronomical Society of New Zealand (RASNZ). In 1977, Frank Bateson published his analysis of the star's light curve in the interval 1953 to 1975, showing a mean interval of 27.33 days between successive outbursts. He drew attention to the two main types of outburst: supermaxima, with a mean brightness of v= 8.64, when the variable remains above

v = 9.5 for longer than 6 days, and normal maxima with a mean maximum of v = 9.45, when the star is brighter that v = 10.5 for 3 days at most (21). Bateson found that supermaxima occur semi-periodically at an interval of 179.35 (+/- 12.1) days, which is referred to as the supercycle length.

This behaviour is typical of members of the SU UMa family dwarf novae. These exhibit superoutbursts which last several times longer than normal outbursts and may be up to a magnitude brighter. During a superoutburst the light curve is characterised by superhumps, modulations which are a few percent longer than the orbital period. They are thought to arise from the interaction of the secondary star orbit with a slowly precessing eccentric accretion disc. The eccentricity of the disc arises because a 3:1 resonance occurs between the secondary star orbit and the motion of matter in the outer accretion disc.

Since Bateson performed his analysis, photometric observations of VW Hyi have accumulated inexorably, largely thanks to the efforts of amateur astronomers. The result is that there is now more than sixty years of photometry available for analysis from the AAVSO International Database, which also incorporates observations from the RASNZ whose members have contributed much of the data (Figure 9). An analysis by the present author shows that around 120 superoutbursts were recorded between 1953 August and 2013 April with a mean supercycle length of 181.4 (+/-17.5) days. Once again the standard deviation of 17.5 days, or about 10% of the supercycle, shows that the ephemeris of superoutbursts is not a very useful predictor of future superoutbursts. As stated before, outbursts of dwarf novae, including superoutbursts, are only quasi-periodic! This is further illustrated by the range of supercycle lengths which were observed, ranging from 135 days to over 220 days (see Figure 10a which shows the number of superoutbursts at different supercycle lengths). However, if one plots the O-C residuals between the observed superoutburst time and time predicted by the ephemeris, there appears to be distinctive and systematic trends, as shown in Figure 10b. There are times when the supercycle appears to be lengthening and times when it appears to be shortening. These trends, which may be related to evolution of the accretion disc and/or changes in the rate of accretion, will be discussed in a later paper which is in preparation and are beyond the scope of this Address. Clearly, VW Hyi, one of the brightest dwarf novae in the sky, still has many secrets to reveal.

What is apparent from this analysis of VW Hyi is that systematic amateur observations of CVs over a long period of time have yielded data which are available for anyone to mine and which may reveal new insights into the behaviours of the system. But the analysis has also revealed a rather worrying trend: in recent years rather fewer visual observations of the star have been made. The consequence of that there is now the very real possibility that future outbursts will not be fully characterised and if the observations become too sparse, it may become impossible to be certain whether an outburst is a normal one or a superoutburst. It is true that more CCD data are available in recent years, but these are often short time series photometry runs, over a few hours, which do not add much to defining the overall profile of the light curve. These trends are by no means unique to VW Hyi. Continued visual monitoring of dwarf novae, especially the so-called "legacy systems" with very long historical runs, will be important in the age of large surveys.

**Searching for dwarf nova outbursts**

Whilst the sky surveys will certainly harvest many outbursts of CV's, there is still a role for the amateur to get there first, which can be important to capture and characterise the outburst in its early stages of development. To this end, one of my own observational programmes is to patrol for outbursts of poorly characterised CVs. These include targets on the VSS Recurrent Objects Programme led by Gary Poyner. This comprises a list of around 100 objects that have rarely (or, in certain cases, never) been observed in outburst, and CV candidates identified by recent surveys such as the SDSS.

Visual observations of CV outbursts continue to be of great value as evidenced by Gary Poyner's work conducted not far from the centre of Birmingham with his 50 cm Dobsonian. However, CCD cameras coupled to modest telescopes have enabled amateurs to probe fainter CVs in recent years. On a transparent night with no moon, the author's 28 cm SCT (Figure 11a) and Starlight Xpress CCD camera can detect stars in outburst at magnitude 18-18.5 in a single 30 second exposure. A 10 cm refractor (Figure 11b) under similar conditions is not far off magnitude 18 in a 60 second exposure.

Having detected an outburst, other observers are alerted by email with the aim of following up with time series photometry which can reveal the nature of the CV's and the class of object it represents. Detecting an outburst, especially of a rarely outburst object, is very exciting – and addictive. I still recall the excitement that Gary Poyner and I experienced when we detected V358 Lyr in outburst more than 40 years since its previous known one! (22)

**Superoutbursts and superhumps**

One area that has received considerable attention from both amateurs and professionals is the study of superoutbursts in SU UMa dwarf novae and their characteristic superhumps which can be up to 0.5 magnitudes in amplitude. Figure 12 shows a superoutburst of V342 Cam and its associated superhumps.

Time series photometry by amateurs equipped with relatively modest CCD-telescope combinations has been a popular activity for many years. Not only can this reveal new scientific insights, but it is also relatively simple to do. In fact, anyone who is capable of taking reasonable images of, say, deep-sky objects is more than capable of studying superoutbursts. Two groups that have consistently encouraged amateurs to conduct this type of photometry, and to publish the fruits of the research are the Center for Backyard Astrophysics (CBA), coordinated by Joe Patterson of Columbia University and Enrique de Miguel of Huelva University, and the VSnet collaboration led by Taichi Kato of Kyoto University.

So what information can be gleaned from observing a superoutburst? Well, measuring the superhump period, $P_{sh}$, immediately gives a reasonable idea of the orbital period, $P_{orb}$, of the system, since there exists an empirical relationship between the two. Moreover, in some systems a separate orbital hump, superimposed on the larger superhump profile, which allows an independent measurement of $P_{orb}$ (for an example, see the inset to Figure 12). $P_{orb}$ can also be accurately measured in eclipsing systems from the interval between successive eclipses. Knowing the superhump period excess, $\varepsilon = (P_{sh}-P_{orb})/P_{orb}$, allows one to estimate the mass ratio of the secondary to the white dwarf primary without recourse to spectroscopy (23) (24) (Figure 13). Already a picture of the binary system begins to emerge.

Often people stop observing a superoutburst after the first few days, once $P_{sh}$ has been established, and the excitement dies down. This is a pity as continued monitoring can reveal further secrets. For example, in most cases the value of $P_{sh}$ changes during an outburst, which typically lasts a couple of weeks, as does the amplitude of the superhumps and sometimes there appears to be a correlation between the two, as well as other gross changes in the light curve. The significance of these is not fully understood and definitely warrants further study. Moreover, although the light curve of different superoutbursts of a particular star is broadly similar, there are subtle variations, the importance of which is also not completely clear.

Another hotly debated topic is what actually initiates a superoutburst. In some SU UMa systems, a normal outburst seems to trigger the subsequent superoutburst. Rather few of these precursor outbursts have been studied in detail mainly due to the practical difficultly of actually catching them in the act – again this is where monitoring and prompt photometry by amateurs can be helpful. We did manage to observe a precursor outburst preceding a superoutburst of V342 Cam (Figure 12) during which we observed orbital humps which gave way to superhumps during the rise to the superoutburst itself (25). Similarly, a precursor outburst of NN Cam (26) was well observed. Precursor outbursts were also found in the *Kepler* satellite light curves of V344 Lyr and V1504 Cyg (Figure 14) (27); moreover, superhumps appeared during the descending branch of the precursor outburst, which supports one of the main models of accretion discs: the thermal-tidal instability (TTI) model. In the future, when a sufficient number of examples are available, a global study of precursor outbursts might be possible which will help to test the validity of the TTI model further.

Thanks to the increase in discoveries of new dwarf novae and detections of outbursts by modern surveys, the number of studied objects has dramatically increased in recent years. However, the proportion of superoutbursts which are *well* observed is decreasing. This trend in observations probably reflects the sheer number of newly discovered objects, which means that on a given night there are many potential targets to choose from. Taichi Kato of the VSnet team has published an annual series of nine papers summarising the results of the team's observations of SU UMa stars over the last few years and has formulated some guidance for observing superoutbursts: (28)

• Single-night time-series observations have limited value (except for classification of the object and the initial detection of superhumps). If there are observations on two nights (preferably consecutive nights), we can determine the superhump period better than 0.2% (1σ error), which is necessary to make a reliable comparison with the orbital period.

• Once you have started to observe an object, stick to the same target for as long as you can during the outburst. In general, fresh outbursts tend to be "over observed", while they become under observed as outbursts progress.

• Even after the superoutburst ends, regularly visit the target and obtain snapshot observations, once or twice a night, since rebrightening episodes can occur

• Early observations are very important in characterising the growth of superhumps, when their period is often evolving rapidly, but also to identify precursor outbursts

Detecting superoutbursts and performing intensive photometry when they occur is something that is especially amenable to observing campaigns, such as those conducted by the CBA, VSnet and the BAA VSS. The aim is to get as near continuous coverage as possible of the superoutburst, which is where observers situated at different longitudes around the world (and not all subject to the same weather system!) can be a real advantage.

For example, a recent VSS campaign (29) by 12 contributors, located in UK, USA, Slovakia and Spain, on the dwarf novae CSS 121005:212625+201948 during 2013-15 revealed that it is a typical SU UMa dwarf nova, but it has one of the shortest supercycles of its class, at 70 days. The superoutbursts are interspersed with three to seven short duration (~2 days) normal outbursts each of which is separated by a mean interval of 11 days, but can be as short as 2 days. Part of the light curve showing three superoutbursts and a multiplicity of normal outbursts is shown in Figure 15. The most intensively studied superoutburst was in 2014 November, which lasted 14 days and had an outburst amplitude of >4.8 magnitudes, reaching magnitude 15.7 at its brightest. Time resolved photometry revealed superhumps with a peak-to-peak amplitude of 0.2 magnitude, later declining to 0.1 magnitude. The superhump period was 127.3 minutes.

Prior to the campaign there had been some speculation that CSS 121005:212625+201948 might be a member of the very very rare sub-group of SU UMa systems known as ER UMa dwarf novae, which have very short supercycles. However, our results ruled this out. This campaign once again demonstrated the value of intensive and co-ordinated monitoring of cataclysmic variables by amateur astronomers possessing relatively simple equipment, complemented with time-resolved photometry at multiple longitudes during outbursts. Being involved in such campaigns is actually great fun and very quickly a sense of community is formed amongst the observers, with shared ownership of the star in question, and often accompanied by a strong proprietorial desire to find out what "our star" is up to!

**WZ Sagittae systems and period bouncers**

Most SU UMa systems have an orbital period of between 2.5 hours down to ~80 mins. A subgroup of SU UMa systems has been recognised, the WZ Sge stars, which have orbital periods at the lower end of the range and have unusually large amplitude (~8 magnitudes) and rare outbursts (years to decades). For a recent review of WZ Sge systems, and further details about classification the reader is directed to reference (30). Because relatively few WZ Sge stars have been observed in outburst, time resolved photometry of their outbursts continues to be of great value. At the beginning of a superoutburst, the light curve usually shows orbital humps, but these soon evolve into superhumps. Then, just when the outburst appears to be over and the system is returning to quiescence, one or more rebrightening events sometimes occur. The record holder is EZ Lyn, which showed 11 rebrightenings during its 2006 outburst – plenty to keep one on one's toes! EZ Lyn was also unusual in going into outburst in 2010 only four years after the previous one, which took the CV community by surprise. This time there were six rebrightening episodes (Figure 16).

Apart from the excitement related to their rare outbursts and exquisite light curves, WZ Sge stars also attract much attention for what they can tell us about the evolution of CVs. The evolution of SU UMa systems is driven by gravitational radiation which results in a loss of angular momentum in the binary. This causes the orbital period to decrease and lowers the mass transfer rate (hence longer intervals between outbursts). As the system continues to evolve, the binary separation continues to decrease. Eventually, the mass of the secondary star becomes so low (< 0.08M$_\odot$) that hydrogen fusion ceases and it becomes degenerate like a white dwarf. At this point the star reaches the period minimum of around 80 mins. However, white dwarfs exhibit a strange effect which is that as they lose mass, their radius *increases*, leading to a slightly longer period. Systems which have evolved beyond the period minimum are therefore referred to as "period bouncers". By this time the system has become very faint: the secondary has a very low mass and the mass transfer rate will decline as the orbital period lengthens. What is left is a planetary-sized brown dwarf orbiting a white dwarf.

Period bouncers are likely to exist among the WZ Sge family, although it is tricky to confirm whether any actually contain the tell-tale brown dwarf and there are no definitively confirmed examples. It has recently been suggested that the southern CV, SSS J122221.7−311525, might be the most highly evolved period bounce candidate, having passed through the period minimum and evolved to a relatively long P$_{orb}$ of ~110 min (31). Other candidate period bouncers include several stars on the VSS Recurrent Objects Programme including , EG Cnc, SDSS J103533.02+055158.3, WZ Sge (32), PQ And, AL Com, EZ Lyn, NSV 24966, SDSS J150137.22+550123.4, UZ Boo and NSV 25966 (33). All these objects warrant careful monitoring and an alert should be triggered as soon as an outburst is detected, for it is in the early stages of the outburst that much information can be gleaned from time resolved photometry. This is something that none of the current surveys can achieve.

Actually, CVs with orbital periods below the period minimum are known. One such is OV Bo. This deeply eclipsing system is unusual as it has a low abundance of metals and it is the only known Population II star among the several thousand known CVs.  In the spring of 2017, the first recorded outburst of OV Boo was followed by amateurs around the world, when it brightened to 11$^{th}$ magnitude from magnitude ~20.

### It's not all about the outbursts! Low states in SW Sex stars and Polars

It might be thought that all the excitement relating to CV observing is related to their outbursts. However, this is not the case, for they show numerous other patterns of behaviour in their light curves and for many systems it is the occurrence of a long anticipated *faint* state that gets the CV community excited.

Let's take the SW Sex stars, for example. These are a sub-class of nova-like CVs originally proposed by Thorstensen et al. in 1991 (34). It initially comprised only eclipsing nova-like stars with typical orbital periods of 3 to 4h and which exhibit single-peaked emission lines, strong He II emission and transient absorption features at orbital phase 0.5 irrespective of the inclination (35), (36). The SW Sex class was later extended to lower orbital inclination non-eclipsing systems which display the same spectroscopic characteristics.

SW Sex stars have high luminosities and hot white dwarfs, implying extremely high accretion rates. (37) Recently it has been suggested that SW Sex stars represent the dominant CV population in the 3 to 4h orbital period range, which, if true, implies that the SW Sex phenomenon is likely to be an evolutionary stage in the life of a CV. (35), (38) This makes the study of SW Sex stars particularly important to our understanding of CV evolution.

Some SW Sex stars show occasional faint states when mass transfer is reduced or even completely stopped, resulting in a drop in brightness of 3−5 magnitudes. The stars can stay at these low levels for weeks or months before rising again to their normal state. This behaviour is very similar to another class of CV, the VY Scl systems. The cause of these faint states is not completely understood, although one suggestion is that it may be due to star spot activity on the secondary: when the star spot moves to the point of the accretion flow, the flow is disrupted. It is therefore important that low states are observed to determine what is happening in the system on the approach to and during the faint state. Approximately one-third of SW Sex stars have been observed to exhibit low states.

HS 0455+8315 was identified as an SW Sex star by a very good friend of the VSS, Boris Gänsicke (Warwick University) and his team (39) during follow-up observations of CV candidates from the Hamburg Quasar Survey. They found that the star was generally 15[th] magnitude, but it undergoes deep eclipses of around 1.5 magnitudes. Subsequent analysis of the times of eclipse revealed $P_{orb}$= 3.569 h. (35) The BAA's David Boyd has studied the eclipse period in great detail and found it to be constant to 1 part in 2 million over an interval of ten years (40).

In a systematic study of the behaviour of HS 0455+8315 over a 14-year interval (41), we found two occasions when the system faded to a faint state at magnitude 19 to 20 for ~500 days (Figure 17a). Low states are particularly sought after as they sometimes allow the components of the binary to be studied without the interference of the bright accretion disc. Our detection of a low state in HS 0455+8315 triggered photometry with the IAC-80 0.82m telescope located at the Observatorio del Teide on Tenerife (Figure 17b). A comparison of eclipse photometry during the normal state and near the minimum showed that the eclipses had very similar profiles and that there were irregular out-of-eclipse modulations with peak-to-peak amplitude up to 0.5 magnitudes. This behaviour is typical of the flickering inherent to accreting CVs. From this we concluded that accretion was in fact still occurring during the minimum. We continue to look out for future low states to see whether accretion actually switches off at some point. Our work identifying high and low states in HS 0455+8315 was sufficient for it to be identified as a VY Scl system in the official "Big List" list of SW Sex stars maintained by Don Hoard of the Max Planck Institute for Astronomy in Heidelberg. (42)

Another category of CV in which high states and low states are observed is the Polars. Here the white dwarf has a very strong magnetic field of tens of millions of Gauss which means that the accretion stream flows along the magnetic field lines down accretion columns which impact violently at the poles (Figure 1b). The light emitted from the accretion columns can account for half the light output of the system and it is highly polarised (hence the name "Polar"). Where the matter in the accretion column hits the pole, hard X-rays are produced. The magnetic field keeps the rotation of the white dwarf synchronised with the orbital period of the secondary star. In other words, the same hemisphere of the white dwarf always faces the donor star.

Since there is no accretion disc, Polars do not have dwarf nova-type outbursts. Instead, they exhibit high states and low states which differ by 2 to 3 magnitudes and they can remain in a particular state for months or years. The timescales on which the system switches from one state into the other also varies considerably, some transitions occur in a few days whereas others occur more gradually over several months. During the low state, mass transfer decreases to a trickle, or may even cease altogether. Again, the exact cause of these mass transfer variations is still unknown, but theories favour stellar activity on the donor star, such as star spots that block the accretion flow. For the last decade, Gary Poyner has managed a monitoring programme, with the encouragement of Boris Gänsicke at Warwick University (43), which has revealed the long-term behaviour of a number of Polars. The results of the first 5 years of the programme (2006 -11) were published in the *Journal* (44). Although the specific Polar programme has recently been wound up, further observation of these systems is encouraged by the VSS.

**Measuring the white dwarf spin periods of Intermediate Polars**

By contrast to the Polars discussed in the previous section, the white dwarfs in CVs known as Intermediate Polars (IPs) have a weaker magnetic field (1 million to 10 million Gauss). Whist the magnetic field is not strong enough to control the accretion flow completely, it is still sufficient to disrupt the inner portion of the accretion disc (Figure 1c). So rather than the material gradually spiralling onto the primary, as in non-magnetic systems, the magnetic field causes it to flow onto the magnetic poles of the white dwarf. Another consequence of the weaker field is that the white dwarf spin is not synchronised with the orbital period. Instead, it rotates about ten times faster than the orbital period, with spin periods in the range of about half a minute to several tens of minutes. The faster spin rate is due to the accretion of high angular momentum material from the secondary.

Measuring white dwarf spin periods of IPs has become a popular sport for amateur photometrists. In many cases it is relatively straightforward to determine the spin period from the small modulations it impresses onto the light curve. The Center for Backyard Astrophysics has specialised in these measurements for a number of years with the aim of determining whether the spin rate of specific systems is changing. The CBA has amassed years, and in some cases decades, of spin rate measurements – keeping time on IPs is definitely a long-term project, where persistence and patience pay off and because of the significant telescope time involved, and the responsibility essentially falls to amateurs.

The curious thing is that whilst many theories predict the white dwarf should "spin up" with time, some actually spin down, while others alternate between spin-up and spin-down. This means the theory is not quite right (the physics involved in the interaction of the accreting plasma and the magnetic field is very complex and difficult to model!) and that further observations are required. Spin-up of the white dwarf is thought to be due to an increase in accretion rate as the torque exerted by the accreting gas increases. By contrast, if the accretion rate goes down, the breaking effect of the magnetic fields causes the white dwarf to spin down. The CBA regularly conducts campaigns on specific IPs to keep tabs on their spin periods.

Amateurs equipped with photometric filters might even be able to shed light on some of the better known IPs. For example, GK Per (45) has a 351-second signal, but it is very difficult to spot in quiescence, which is 95% of the time. Therefore no one has determined its long-term ephemeris. This is important because it has by far the highest accretion rate among all IPs and thus should spin up quickly.  The reason for this failure hitherto is that there is a subgiant of spectral type K in the binary, which overwhelms the light from the white dwarf. This can be subdued with UV photometry, i.e. using a CCD with sufficient UV sensitivity along with a UV filter (46).

A further interesting aspect of IPs, due to the fact that they have an accretion disc, albeit truncated, is that they can undergo dwarf nova outbursts. Several IPs have also shown occasional, very short, outbursts which might be related to episodes of enhanced accretion rate from the secondary, although a recent paper suggests that coupling of the magnetic fields generated by a magneto-rotational instability with the magnetic field of the white dwarf can, under particular conditions, generate an instability (47). Since these brief outbursts might only last 1-2 days, or even only a few hours, amateur detection and follow-up continues to be important. Examples, although yet to be conclusively identified as IPs, where the VSS has coordinated observing campaigns include V1316 Cyg, HW Boo and, 1RXS J140429.5+172352 and their outburst characteristics are shown in Table 2.

Finally, some IPs have shown occasional low states, although not as frequently as in Polars. The reason for these is not known, but it might be related to X-ray irradiation of the secondary. The 2016 low state of FO Aqr was found to be associated with complete disappearance of the accretion disc (48). Finding an IP in a low state presents the important possibility of observing the white dwarf and the secondary independently – there are very few examples where an IP secondary has been convincingly detected (exceptions are GK Per and DO Dra).

The study of IPs attracts much professional attention at present and more observational data are required to support ongoing modelling work to understand these intriguing systems. In terms of the range of their optical behaviour, there is plenty to look out for! Koji Mukai maintains an online list of conformed and possible IPs on his NASA website. (49)

**Novae and Recurrent novae**

One field, which was formerly the domain of amateur astronomers, but where the surveys have already made a major impact, is the discovery of novae. The BAA's George Alcock (1912 – 2000) remains famous for his discovery of five novae, but nowadays the discovery of a nova by an amateur is uncommon. Nevertheless, amateurs continue to patrol for these new stars, not only in our own galaxy, but also further afield. Bromsgrove-based BAA member George Carey has a programme to search for novae in our neighbouring galaxies and in 2015 he co-discovered one in M31 using his homebuilt 20 cm reflector equipped with a CCD camera and then a second in 2017 September (Figure 18). The latter, M31N 2017-09b, is a classical nova of Fe II spectroscopic class (50).

Long gone are the days when one might accidently stumble across a nova in the course of other observational activities such as occurred to the young Tony Ellis of Llandudno Junction in 1942 (51). The news was announced on a BAA Circular issued on November 14 that Ellis, an inshore fisherman, had discovered a nova in Puppis the previous day. In a letter penned by Ellis, he described how he had been using his 4-inch (10 cm) refractor (Figure 19) to observe star clusters in Monoceros and Puppis on the morning of November 13 when he noticed the third magnitude object at 04.25 UT. At a declination of -35° 10' 36" (52), the object must have been very low down in the sky from his observing location, although he noted that conditions were excellent at his site near Bryn Pydew 400 ft (120 m) above sea level and he had "an unobstructed view to within 1° of the true horizon" (53). Even with elevation and atmospheric refraction helping to lift the object a little, the object cannot have culminated much more than 1.5 to 2° above the horizon (54), making this an impressive feat to not only spot it, but recognise it as an unfamiliar object at that altitude above the horizon, but it certainly seems quite possible. Ellis telephoned the Greenwich Observatory to report his discovery. The Journal subsequently carried reports by Will Hay, who was able to observe it with his naked eye on the morning of November 14 at 04.15UT, and by Henry Wildey, observing from Parliament Hill, Hampstead on November 17. Hay saw the nova again on November 22 and, although it had faded, it was easily seen in binoculars. By November 24, he picked it up in his 3½-inch (8.9 cm) Cooke refractor at fifth magnitude and noted that it was reddish in colour.

The British Pathé newsreels carried a feature on Ellis which was presented with its usual enthusiasm and breezy commentary. However, unbeknown to Ellis, because of the poor communications as a consequence of the War, it transpired that others had spotted the nova before him. The first was Bernhard H. Dawson, a young American living in the city of La Plata, Argentina, who had spotted the nova on November 9. Dawson was employed at the city's observatory, but when he made his observation, he had finished work and had gone up onto the roof of his residence to scan the skies with his naked eyes. At the time, the object was about magnitude 1.5. Two days later, Nova Puppis, now known as CP Pup, reached its maximum brightness of magnitude +0.3 and was one of the brightest novae of the twentieth century. By the time Ellis picked up the object, it was already in rapid decline. Soon after the discovery, Ellis was put forward for election to the BAA (55) and the RAS.

It is possible that all novae eventually recur on some timescale and there are a few that have been observed to erupt more than once – these are the recurrent novae. The galactic recurrent novae are a rather select group with only ten confirmed members (Table 3). The short recurrence period is driven by a combination of a high mass white dwarf and a high accretion rate. One such is T Coronae Borealis, which has undergone two outbursts: in 1866 and 1946. The first was detected by the Irish astronomer John Birmingham (1816–1884) on 1866 May 12, at magnitude 2.0, about 10 magnitudes above its normal quiescent level. The situation surrounding the second outburst will be familiar to anyone who has read the wonderful autobiography of the American amateur astronomer, Leslie C. Peltier (1900-1980), *Starlight Nights* (if you have not read this book, then I thoroughly recommend you to do so at the earliest opportunity: it is a brilliant and sympathetic description of why so many of us love observing the night sky). T CrB had been on Peltier's observing list for 25 years before it decided to go into outburst on 1946 February 9. But when Peltier's alarm sounded at 2.30 on a clear but freezing morning, he felt he had a cold coming on and that it would be advisable to stay in bed. He thus missed its dramatic reappearance on the stage after 80 years, noting: "I alone am to blame for being remiss in my duties, nevertheless, I still have the feeling that T could have shown me more consideration. We had been friends for many years; on thousands of nights I had watched over it as it slept, and then it arose in my hour of weakness as I nodded at my post. I still am watching it but now it is with a wary eye. There is no warmth between us anymore." (56) I know I have experienced similar feelings on winter mornings and, with a sense of guilt, decided to remain warm and snug in my bed!

It is now more than 70 years since the last episode, so could T CrB be about to make another appearance? Well, during 2015 it entered a "super-active state" characterised by an increase in the mean brightness (~0.7 mag) and a bluer colour (57). John Toone of the VSS has detected a brightening trend since April 2015 in his visual observations (58) (Figure 20). Intriguingly a super-active state was recorded in 1938, just 8 years before the last eruption. Only time will tell when the next eruption will occur – and it will be interesting to see whether it is an amateur astronomer that makes the detection, considering how bright the object is, or one of the sky surveys. Amateurs certainly have a distinct advantage at the start and end of observing seasons as they can follow the object well into the twilight.

Let's turn now to another recurrent nova, this time outside our galaxy. The remarkable recurrent nova M31N 2008-12a, within the Andromeda Galaxy, has been observed in eruption 12 times in the last few years with a recurrence period of around 350 days (59). This is much shorter than the next shortest recurrence period of a nova, U Scorpii, which went off twice in eight years (1979 and 1987). During 2016-17, Dr. Matt Darnley (Liverpool John Moores University) and Dr Martin Henze (European Space Astronomy Centre, Spain) organised an observing campaign amongst professionals and amateurs to detect its next eruption. The present author was involved in the campaign, which was not for the faint-hearted as the nova only reaches magnitude 18.3 - 18.5 at its brightest and the outburst only lasts a few days. This was a very exciting campaign and anticipation mounted as the next outburst during 2016 September loomed. However, as the autumn marched on without any sign of an eruption, nerves were becoming frayed! There was the occasional false alarm, but members of the team were soon able to weed these out. Then, in mid-November Matt Darnley pointed out that on December 15 we would reach the 3-sigma confidence level (99.73%) of the eruption ephemeris. Was something wrong? Had there been a miscalculation? Well, in the end the long-awaited outburst finally occurred on December 12 (60), much to everyone's relief! This was 444 days after the previous one – much longer than any previous eruption interval.

So when can we expect the next eruption? If we treat the 2016 eruption as an outlier and assume a mean recurrence period of ~348d with a standard deviation of ~30d (from the 2008-2015 statistics), this takes us to late November 2017 with a statistical 1-sigma uncertainty of about one month.

Understanding the nature of M31N 2008-12a is important as the white dwarf is very close to the Chandrasekhar limit, making it the leading pre-explosion supernova Type 1a progenitor candidate. The campaign continues to look out for future outbursts. Amateur observations during the twilight periods at the beginning and end of the observing season are particularly valuable as other surveys do not operate during these times.

**Spectroscopy: a new tool for the amateur nova enthusiast**

The availability of relatively low cost spectroscopic equipment is stimulating a surge of interest in spectroscopy with amateur-sized telescopes. This is something the BAA has also helped to encourage through the provision of Ridley grants to help Members purchase the equipment, as well as by providing training in its use. In general, CVs are not a major focus for amateur spectroscopists as our spectra don't shed much new light on their behaviour and most uncharacterised CVs are too faint for our instruments. Nevertheless, there are some specific cases such a V Sge, classified as a "nova-like" CV but a rather enigmatic object, where amateur spectra are providing new information.

On the other hand, one area where spectroscopists with amateur-sized telescopes are beginning to play an important role is in the confirmation of nova and dwarf nova discoveries. BAA member, Paul Luckas observing at Perth, Western Australia used a 35 cm telescope equipped with an Alpy 600 spectrograph (Figure 21) to confirm two southern novae in the same week during 2016 September (61) (62). The first of these (V5855 Sgr or Nova Sgr 2016 number 3) was announced as a potential nova by the Central Bureau of Astronomical telegrams on 2016 Oct  20.383. Paul picked it up about 4 hours later and took the first spectra (Figure 22). Time was of the essence as by this time the target was setting rapidly. The spectrum was therefore obtained under rather challenging conditions, but it provided essential confirmation that it was indeed a classical nova caught whilst in the early optically thick "fireball" stage. There were prominent Balmer lines as well as Fe and the P Cygni profiles which was indicative of outflowing material.

One of the brighter novae of recent times was Nova Del 2013. Following its discovery by Japanese amateur Koichi Itagaki on 2013 August 14, there was global amateur global spectroscopy campaign to follow its decline. Similar follow-up work on other novae by amateurs has been conducted, notably a programme supported by Professor Steven Shore of Pisa University, and this is turning into a fruitful area of pro-am collaboration.

Bright dwarf novae also fall within range of amateur spectroscopists. Spectra obtained by two amateurs, Paolo Berardi and Umberto Sollecchia, confirmed the dwarf nova identity of a newly discovered 12[th] magnitude optical transient in Lyra, TCP J18154219+3515598, in 2017 June. Berardi used a 0.23m SCT telescope and Lhires III spectrograph configured for low resolving power (4200-7400Å, 10 Å resolution). Sollecchia used a 0.20m SCT telescope and Alpy 600 spectrograph (63).

**The helium dwarf novae**

Another very rare class of CV is the helium dwarf novae, or AM CVn systems. These are also accreting binaries containing a white dwarf, but the secondary is a helium white dwarf, or a low-mass helium star, or a highly evolved main-sequence star (64). These ultra-compact binaries have very short orbital periods of 5 to 65 min. By contrast to their hydrogen-containing cousins, only a few dozen helium dwarf novae are currently known. With the advent of sky surveys, the number is increasing rapidly, but further long-term studies are required to help characterise their behaviour (Figure 23).

There are in fact three groups of helium dwarf novae - those in a permanently bight state, those in a permanently faint state and those that undergo outbursts – and which group a star is in is related to the orbital period. Once again, amateur astronomers with small telescopes equipped with CCD can do useful work by conducting time resolved photometry when an outburst is detected. One of the AM CVn systems a group of amateurs studied is V744 And (65) during its first confirmed superoutburst in 2009 which lasted more than 75 days (Figure 24). Our data revealed for the first time tiny superhumps (0.06 magnitude peak-to-peak) similar to those seen in SU UMa systems – and with the same origin. We also observed six echo outbursts leading us to suggest it is a helium analogue of the WZ Sge system. Detailed analysis allowed us to estimate its orbital period, which had hitherto been unknown, as 37 minutes. Noting the similarity of its orbital period to another AM CVn system, SDSS J124058.03-015919.2, prompted us to wonder whether this too could be an outbursting system. Sure enough, interrogation of the online All Sky Automated Survey (ASAS-3) and Catalina Real-Time Sky Survey (CRTS) databases revealed it had been in outburst in 2005. This again illustrates the value of data-mining surveys and that surveys can be the friend of amateur astronomers, not the foe!

Comparing the properties of helium and hydrogen accretion discs may provide a better understanding of accretion disc physics. AM CVn stars also hold great interest as mass transfer in these systems is believed to be driven exclusively by gravitational radiation. Understanding their population may help in the interpretation of gravitational wave detections by the new generation of gravitational wave observatories.

The harvest of AM CVn stars from the new surveys is expected to be bountiful. A recent discovery gives a hint of the exciting times ahead. Gaia14aae, located about 730 light years away in Draco, was observed in outburst by an international team of both professionals and amateurs led by Dr Heather Campbell of Cambridge University's Institute of Astronomy (66). They found an orbital period of 49.7 min. Actually three outbursts were detected during 4 months; it was rather surprising that Gaia14aae shows outbursts at all, because a system with such a long orbital period is expected to have a stable, cool disc. Moreover, it was found to be only the third eclipsing AM CVn star known, and the first in which the WD is totally eclipsed (Figure 25). The deep eclipses were first detected via amateur CCD photometry and it is anticipated that future observations of Gaia14aae have the potential to lead to the most precise determinations of the parameters of the constituent stars of any AM CVn system discovered to date.

As Dr. Campbell noted, "It's really cool that the first time that one of these systems was discovered to have one star completely eclipsing the other, that it was amateur astronomers who made the discovery and alerted us. This really highlights the vital contribution that amateur astronomers make to cutting edge scientific research." (67)

**Black hole astronomy from your back garden? The story of V404 Cyg**

Until now, we have only considered binary systems which contain an accreting white dwarf. However, there is an even more exotic class of binary system in which the accretor is a neutron star or even a black hole. These systems are strong X-ray sources and are sometimes called Low Mass X-Ray Binaries. Some systems also undergo outbursts similar to dwarf novae in which the accretion disc brightens and even shows superhumps. One system is V404 Cyg, which is believed to comprise a 17 $M_\odot$ black hole with a red giant in a 6.5-day orbit around it. Several outbursts of V404 Cyg have been observed: in 1938, 1956 and 1989, when it was identified as the optical counterpart of an X-ray transient detected by the Ginga satellite.

V404 Cyg has been on the VSS Recurrent Objects Programme for many years. When I last spoke about this star during my 2013 George Alcock Memorial Lecture, I mentioned that I monitor it every clear night, hoping that one day when I download an image of the field, the 18th magnitude star at its position will once again be replaced by a beacon shining at 11[th] magnitude! Well, we didn't have too long to wait for the next outburst as the tell-tale burst of gamma rays were detected by the Burst Alert Telescope on NASA's *Swift* satellite on 2015 June 15.77. The outburst was independently detected by Belgian amateur astronomer Eddy Muyllaert in an image he took with the remotely-operated Bradford Robotic Telescope only 9.5 hours later.

The *Swift* detection triggered alerts throughout networks of professional and amateur astronomers around the world, causing hundreds of instruments to point towards it. At its peak, V404 Cyg was the brightest object in the X-ray sky – up to fifty times brighter than the Crab Nebula. A paper in one of the world's most prestigious journals, *Nature*, on V404 Cyg included multicolour photometric observations by the BAA's Ian Miller, Nick James and Roger Pickard, as well as many other amateur observers (68). Remarkably, these observations revealed oscillations on timescales of 100 seconds to 2.5 hours and are thought to be associated with physical processes in the inner accretion disk. Other observers reported very rapid (sub-second) optical pulses (69) which may be associated with jets spewing out some of the accreted material away from the black hole (Figure 26). These jets are believed to be unionised hydrogen and helium ejected at ~1% of the speed of light allowing the material to escape from the gravitational field around the black hole. After leaving the disk, the ejected material expanded and cooled, and was observed to form nebula.

When all the science is said and done, there is still something very special about being able to go out to one's telescope at night and observe the goings on of a black hole! The VSS Recurrent Objects Programme contains another X-ray nova and black hole candidate, V518 Persei – definitely worth keeping an eye on!

**Participating in the new golden age of cataclysmic variable star astronomy**

I hope that this presidential address has given at least some indication of the ways amateurs can contribute to CVs astronomy, although I fear I have only just scratched the surface with the examples I have provided. Furthermore, I hope that I have shown that the emerging sky surveys, far from making amateurs redundant, will actually give us more work to do. We might have to modify our programmes a little, but there is certainly no chance of relaxing! I also hope to have shown that this work can be accomplished by amateurs equipped wide a range of instrumentation, from visual observations with a simple Dobsonian telescope, right through to the most sophistic computer controlled instrument supplemented with a CCD camera or a spectroscope. It caters for all levels of dedication from those who might like to adopt a handful of CVs, which they weave into their other observing programmes, though to the dedicated CV enthusiast. Some may enjoy following the ever-changing ups and downs of well-known systems, some may enjoy the thrill of the chase to spot a rarely outbursting object, whilst others again get satisfaction from knowing that their observations contribute to the characterisation of a newly discovered system.

There is also room for those who do not want to observe at all: there is plenty of scope in data mining exiting CV databases, as the VW Hyi example illustrates. When all is said and done, we must not forget that amateur astronomy is a hobby and the huge advantage we amatuers have as non-paid astronomers is we can chose what to observe and when – purely for our own enjoyment! It's important that amateurs should never feel that feel that they are doing 'work' for professionals, they have to benefit from the collaboration too, guided by their own interests

All these things make for CV astronomy being a very rewarding and absorbing activity. However, if I might speak personally for a moment, for me the most enjoyable aspect of this line of work that observing CVs is an activity that is ideally suited to cooperation and teamwork. There is a real spirit of community amongst CV astronomers which sees no divide between the professionals and amateurs. And it's this social side that gives me more pleasure than anything else, for through this work I have developed many long-lasting friendships. I do hope that others might wish to join this friendly community of CV enthusiasts. Beware, though: the lure of the CVs is strong and can sometimes become an overwhelming passion!

**Acknowledgements**

I wish to thank Elmé Breedt (University of Cambridge), Gary Poyner (BAA-VSS) and Chris Lloyd (University of Sussex), for constructive comments on a draft of this Presidential Address. In addition, I thank David Boyd, Robin Leadbeater, Paul Luckas, Poshak Gandhi (University of Southampton) for helpful discussions. Bill Gray of Project Pluto (publisher of the Guide astronomical software, https://www.projectpluto.com/) provided advice on the observability of Nova Puppis in 1942. Graeme and Louise Watt of Tasmania generously provided material in connexion with the discovery of this nova by Louise's father, Tony Ellis.

Dr Andrew Beardmore, Department of Physics and Astronomy, University of Leicester, kindly gave permission to reproduce his beautiful and informative graphics of cataclysmic variable star systems in this paper, as well as to use his animations of these systems in the Address I gave at Burlington House.

This research made use of the data from the BAA VSS, the AAVSO and the Center for Backyard Astrophysics. I thank all observers who have contributed their observations to these organisations. In addition, I used the NASA/Smithsonian Astrophysics Data System, the AAVSO Variable Star Index, and the SIMBAD Astronomical Database operated through the Centre de Données Astronomiques (Strasbourg, France).

Finally, I thank all the astronomers from around the world, both amateur and professional, with whom it has been my privilege to cooperate on various campaigns since I began observing cataclysmic variables in earnest in 2004. You have been generous with your hard-earned data, with your time and, above all, with your friendship.

**References**

1. *The VSS database may be accessed online via the BAA website at: http://www.britastro.org/vssdb/.*

2. *Warner B., Cataclysmic Variable Stars (Cambridge Astrophysics Ser. 28), publ. Cambridge University Press, Cambridge (1995).*

3. *Hellier C., Cataclysmic Variable Stars - How and Why they Vary, publ. Springer Praxis Books, London (2001).*

4. *Historically, because CVs were observed photometrically and without apparently following any regular pattern, they were referred to as "cataclysmic" (from the Greek word kataklysmos, meaning storm or flood).*

5. *A paper describing Brook's life and work as VSS Director, and his work on SS Cyg, may be read in: Shears J., JBAA, 122, 17-20 (2012).*

6. *Brook C. L., JBAA, 21, 255 (1911).*

7. *Brook C. L., J. Brit. Astron. Assoc., 24, 192 (1914).*

8. *Brook C. L., JBAA, 36, 205 (1926).*

9. *Cannizzo J. K. and Mattei J.A., ApJ. 401, 642-653 (1992).*

10. *Miller-Jones J. C. A., Sivakoff G.R., Knigge C. et al., Science, 340, 950-952 (2013).*

11. *A series of workshops entitled "The Golden Age of Cataclysmic Variables and Related Objects" have been been held in 2011, 2013 and 2015.*

12. *Walker M.F., PASP, 66, 230 (1954).*

13. *Mumford G.S., PASP, 79, 283 (1967).*

14. *Szkody P. and Gänsicke B. JAAVSO, 40, 563-571 (2012).*

15. *Cannizzo J.K. & Mattei J.A., ApJ, 505, 344 (1998).*

16. *Wheatley P.J., Mauche C.W. & Mattei J.A., MNRAS, 345, 49-61 (2003).*

17. *Shears J., Pickard R. & Poyner G., J. Br. Astron. Assoc., 117, 22-24 (2007).*

18. *Shears J., Boyd D. and Poyner G., J. Br. Astron. Assoc. 116, 244-247 (2006).*

19. *Ridgway S.T. et al., ApJ, 796, 53 (2014).*

20. *Luyten W.J., AN, 245, 211 (1932).*

21. *Bateson F.M., NZ Jounal of Science, 20, 73-122 (1977).*

22. *Shears J. et al., J. Br. Astron. Assoc. 120, 43-48 (2010).*

23. *Patterson J., PASP, 113, 736-747 (2001).*

24. *Patterson J., et al., PASP, 117, 1204 (2005).*

25. *Shears J.H. et al., New Astronomy, 16, 311–316 (2011).*

26. *Shears J. et al., J. Br. Astron. Assoc. 121, 6, 355-362 (2011).*

27. *Osaki Y. & Kato T., PASJ, 66, 15 (2014).*

28. *Kato T., PASP, 68, 65 (2016).*

29. *Shears J. et al., J. Br. Astron. Assoc., 126, 178-184 (2016).*

30. *Kato T., PASJ, 67. 108 (2015).*

31. *NeustroevV.V. et al., MNRAS preprint (2017) available at https://arxiv.org/abs/1701.03134.*

32. *A recent paper suggests that the orbital period of WZ Sge is still decreasing, which would mean it is still evolving towards the period minimum, rather than being a bouncer. See Han Z.-T. et al., https://arxiv.org/abs/1705.03155.*

33. *Joe Patterson presented a list of 20 period bouncer candidates in: Patterson J., MNRAS, 411, 2695-2716 (2011).*

34. *Thorstensen J. R. et al., AJ 102, 272 (1991).*

35. *Rodríguez−Gil P., Gänsicke B. T., Hagen H.−J. et al., MNRAS 377, 1747−1762 (2007).*

36. *Rodríguez−Gil P., Schmidtobreick L. & Gänsicke B. T., MNRAS, 374, 1359 (2007).*

37. *Townsley D. M. & Gänsicke B. T., ApJ, 693, 1007 (2009).*

38. *Schmidtobreick L., Rodríguez−Gil P. & Gaensicke B. T., Mem. S.A.It., 83, 610−613 (2012).*

39. *Gänsicke B.T. et al., The physics of cataclysmic variables and related objects, ASP Conference Series, 261, 623−624 (2002).*

40. *Boyd D., JAAVSO, 40, 295−314 (2012).*

41. *Shears J., JBAA, 126, 42-46 (2016).*

42. *Hoard D. W., http://www.dwhoard.com/biglist. The constitution of the Big List is described in Hoard D. W. et al., AJ, 126, 2473 (2003).*

43. *Gänsicke B., VSSC, 129, 7-9 (2006).*

44. *Poyner G., J. Br. Astron. Assoc. 123, 108-114 (2013).*

45. *GK Per = Nova Aquilae 1918.*

46. *Patterson J., posted on CBA-news group (2017 October 1).*

47. *Hameury J.-M. & Lasota J.-P., submitted to A&A (2017) available at https://arxiv.org/abs/1703.03563.*

48. *Hameury J.-M. and Lasota J.-P., Accepted for publication in A&A (2017). Pre-print available at https://arxiv.org/abs/1707.00540.*

49. *http://asd.gsfc.nasa.gov/Koji.Mukai/iphome/catalog/alpha.html.*

50. *Williams S.C. & Darnley M.J, ATel #10754 (2017).*

51. *Ellis was known as Tony to his family and friends, but in official documents he was Gordon.*

52. *The declination quoted is the J2000.0 declination precessed to 1942 Nov, with thanks to Bill Gray of Project Pluto.*

53. *"The Nova in Puppis", JBAA 53, 42-43 (1942).*

54. *According to Bill Gray of Project Pluto, the calculated altitude at the time of the discovery was 1.23° without altmospheric refraction. Assuming a "standard" atmosphere, temperature 10°C, relative humidity 20%, one might expect a refraction of about.* 0.35°, bringing the object up to an altitude of 1.58°. The altitude of the observer might have raised that slightly further assuming a clear southern horizon.

55. *Ellis was elected BAA member on 1943 March 31, proposed by Vice President F.J. Sellers.*

56. *Peltier L.C., Starlight Nights: the adventures of a stargazer, publ. Macmillan London & Edinburgh (1967).*

57. *Munari U., Dallaporta S. and Cherini G., New Astronomy, 47, 7-15 (2016).*

58. *Toone J., BAA VSS Circular 169, 6-9 (2016).*

59. *Darnley M.J. et al., ApJ, 833, 38 (2016).*

60. *Itagaki K. et al., ATel 9848 (2016) .*

61. *Luckas P., ATel #9658 (2016) available at http://www.astronomerstelegram.org/?read=9658.*

62. *Luckas P., ATel #9678 (2016) available at http://www.astronomerstelegram.org/?read=9678.*

63. *CBAT "Transient Object Followup Reports", TCP J18154219+3515598, http://www.cbat.eps.harvard.edu/unconf/followups/J18154219+3515598.html.*

64. *Nelemans G., 'The Astrophysics of Cataclysmic Variables and Related Objects' in Proceedings of ASP Conference Vol. 330, J.-M. Hameury & J.-P. Lasota, San Francisco, ASP (2005).*

65. *Shears J. et al., JBAA, 122, 49-53 (2012).*

66. *Campbell H.C. et al., MNRAS, 452, 1060-1067 (2015).*

67. *Cambridge University online article "Gaia satellite and amateur astronomers spot one in a billion star" at http://www.cam.ac.uk/research/news/gaia-satellite-and-amateur-astronomers-spot-one-in-a-billion-star.*

68. *Kimura M. et al., Nature, 529, 54-58 (2016).*

69. *Gandhi P. et al., MNRAS, 463, 1822-1830 (2016).*

70. *Shears J. et al., J. Br. Astron. Assoc. 125 (4), 236-241 (2015).*

71. *Shears J. et al., . Assoc. 125 (6), 358-362 (2015).*

72. *Schaefer B.E., "Comprehensive Photometric Histories of All Known Galactic Recurrent Novae" in AAS Meeting 213, id.491.04; Bulletin of the American Astronomical Society, 41, 467 (2009).*

73. *The nebula was discovered on 1852 October 11 by John Russell Hind, who also discovered U Gem as mentioned elsewhere in the paper.*

74. *Osaki Y. & Kato T., PASP, 65, article id 50 (2013).*

75. *Breedt E., "An update on the AM CVn family" in Proceedings of Science, The Golden Age of Cataclysmic Variables and Related Objects - III (2015).*

| Timescale | Phenomenon |
|-----------|-----------|
| Minutes | Flickering in dwarf novae, orbital periods of AM CVn systems |
| Hours | Orbital periods of novae, dwarf novae; rise time of dwarf nova outburst |
| Days | Normal outburst lengths of dwarf novae |
| Weeks | Outburst length of superoutbursts in short orbital period dwarf novae; outburst recurrence time of normal outbursts in short orbital period dwarf novae |
| Months | Outburst recurrence time of longer period dwarf novae; various state changes in polars; declines in novae |
| Years | Outburst recurrence timescales of the shortest period dwarf novae and the recurrence times in recurrent novae |

Table 1: Timescale of variability in cataclysmic variables

| Star | Typical quiescence mag | Outburst duration (d) | Outburst amplitude (mag) | Reference |
|------|------------------------|-----------------------|--------------------------|-----------|
| V1316 Cyg | 17.4 | 1-2 | ~1.4 | (18) |
| HW Boo | 16.8 | 2-5 | ~3 | (70) |
| 1RXS J140429.5+172352 | 17.2 | 3-5 | ~4 | (71) |

Table 2: Suspected IPs showing short-duration outbursts

| Name | Discoverer | Magnitude range | Confirmed outburst years |
|---|---|---|---|
| CI Aquilae | K. Reinmuth | 8.6-16.3 | 2000, 1941, 1917 |
| V394 Coronae Australis | L.E. Erro | 7.2-19.7 | 1987, 1949 |
| T Coronae Borealis | J. Birmingham | 2.5–10.8 | 1946, 1866 |
| IM Normae | I.E. Woods | 8.5-18.5 | 2002, 1920 |
| RS Ophiuchi | W. Fleming | 4.8–11 | 2006, 1985, 1967, 1958, 1933, 1898 |
| V4287 Ophiuchi | K.Takamizawa | 9.5-17.5 | 1998, 1900 |
| T Pyxidis | H. Leavitt | 6.4–15.5 | 2011, 1967, 1944, 1920, 1902, 1890 |
| V3890 Sagittarii | H. Dinerstein | 8.1-18.4 | 1990, 1962 |
| U Scorpii | N.R. Pogson | 7.5–17.6 | 2010, 1999, 1987, 1979, 1936, 1917, 1906, 1863 |
| V745 Scorpii | L. Plaut | 9.4-19.3 | 2014, 1989, 1937 |

Table 3: Confirmed galactic recurrent novae from reference (72)

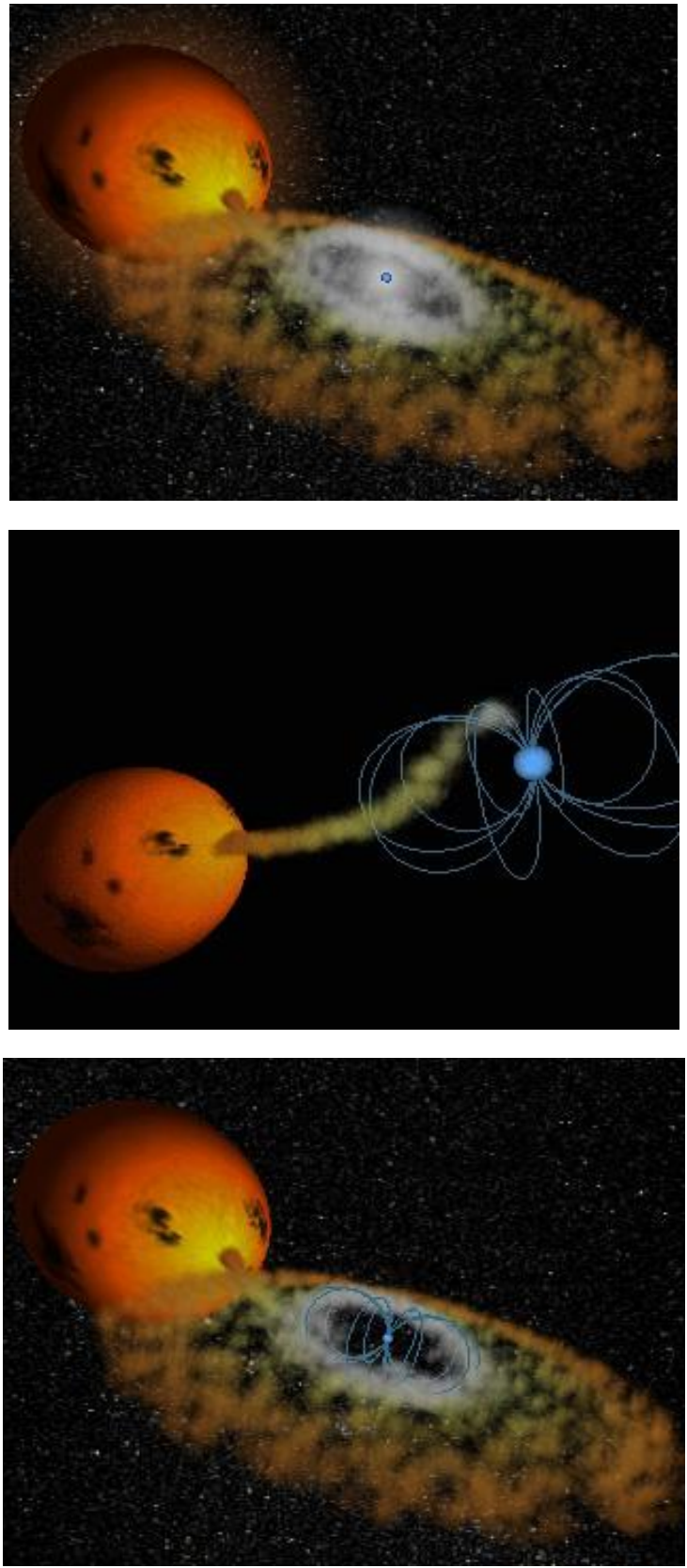

Figure 1: Three types of cataclysmic variable. (a) Non-magnetic CV, (b) Magnetic CV or "Polar" and (c) Intermediate Polar (images courtesy Dr Andrew Beardmore)

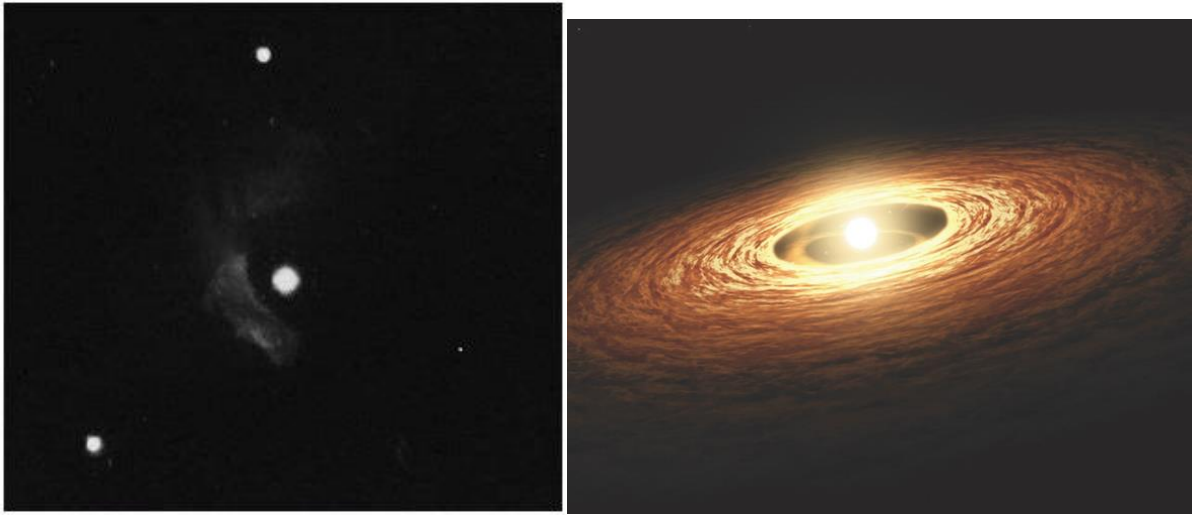

Figure 2. (a) T Tauri showing the nearby Hind's Variable Nebula (73), NGC 1555. drawing by Gary Poyner, 2004 December, 35cm SCT x290. (b) Artist's impression of a T Tauri star surrounded by clouds of gas and dust. (NASA)

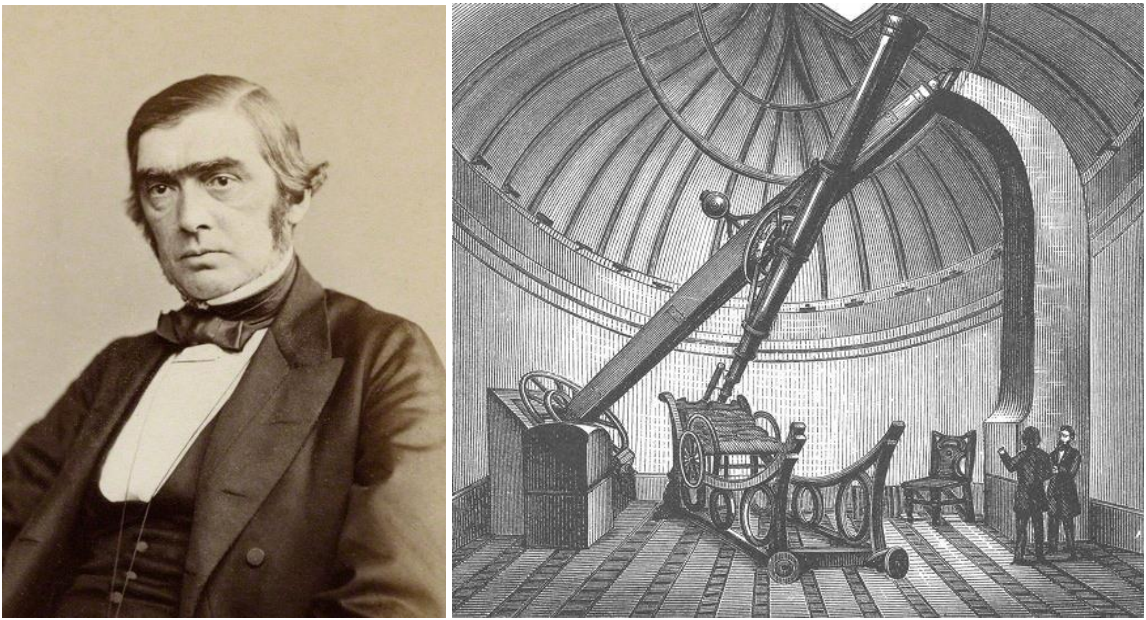

Figure 3: (a) John Russell Hind ca. 1860s, (b) George Bishop's observatory in Regent's Park, 1850

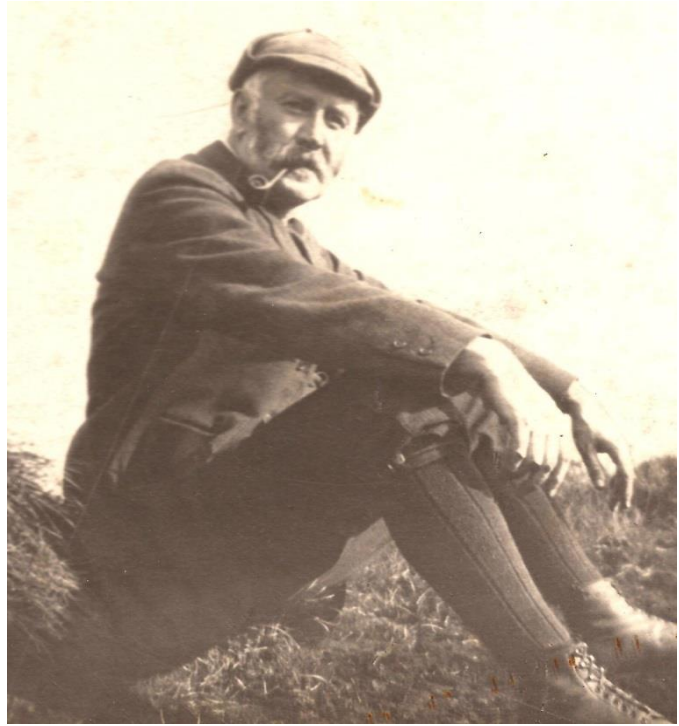

Figure 4: Charles Lewis Brook (1855−1939)

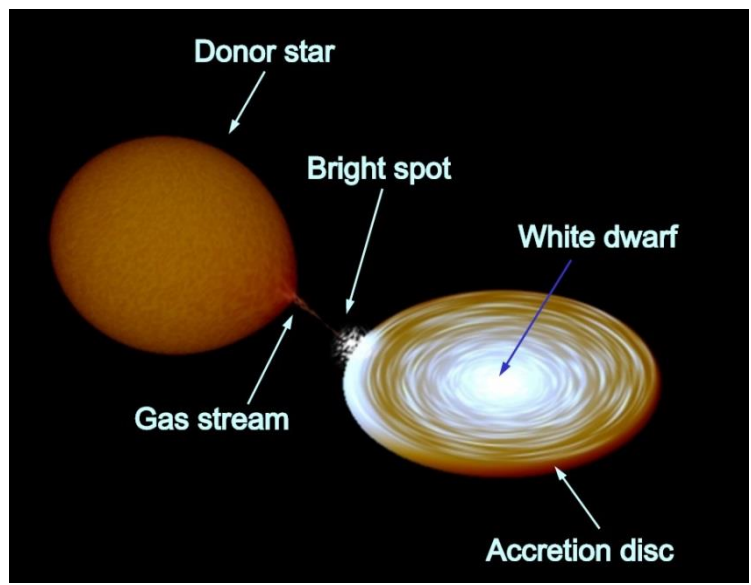

Figure 5: the anatomy of a CV

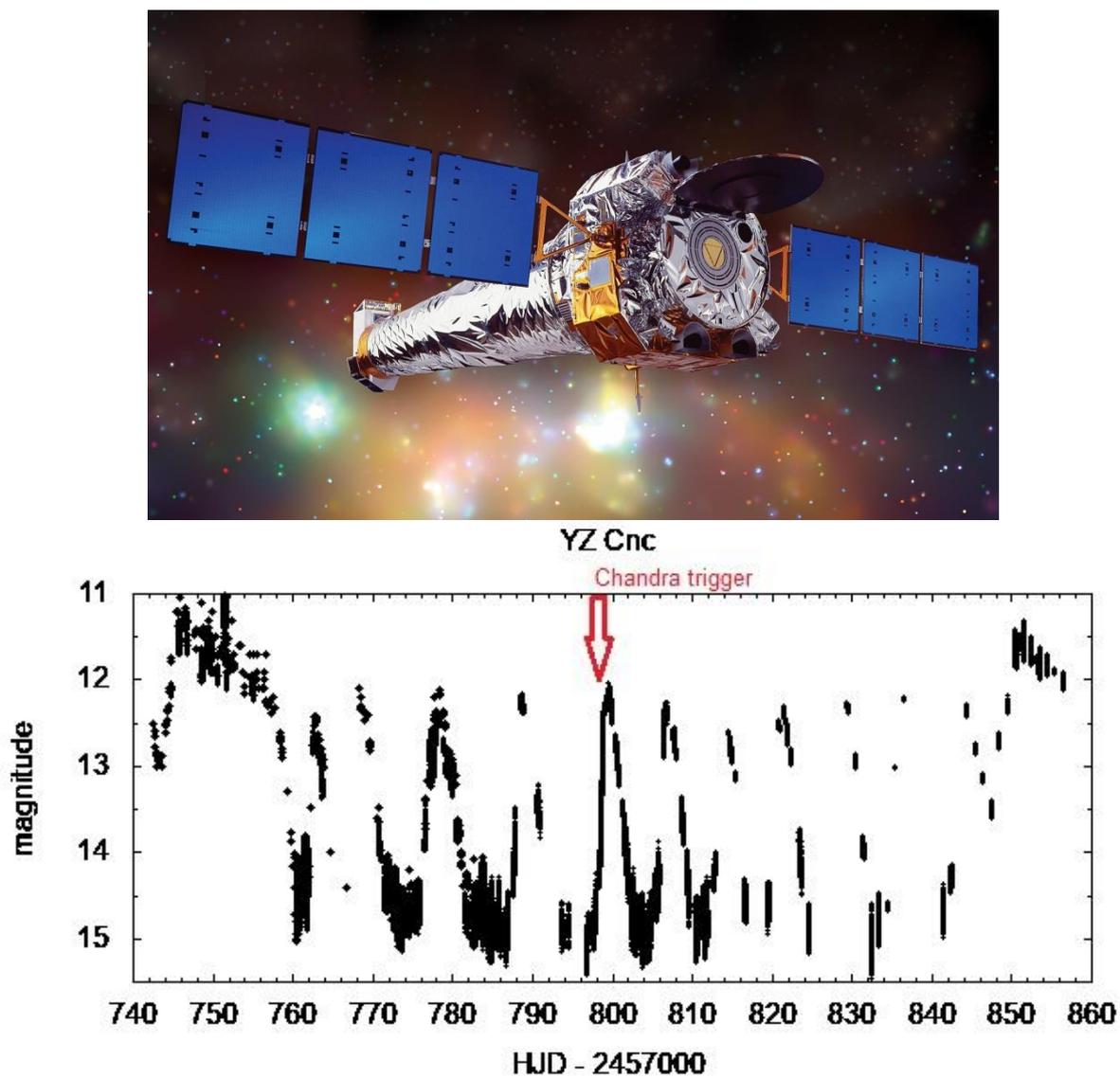

Figure 6. Top: Artist's impression of the Chandra X-ray Observatory (image: NASA). Bottom: Plot of observations of YZ Cnc starting at HJD 2457740 (2016 Dec 17) contributed to the Center for Backyard Astrophysics and AAVSO databases (image courtesy of Dr Enrique Miguel, Huelva University, Spain)

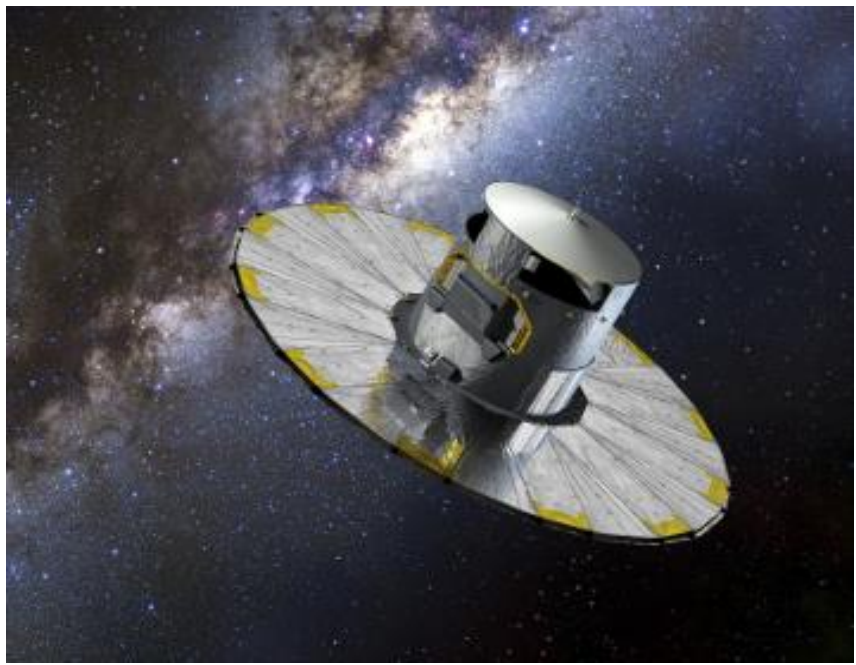

Figure 7 (a): Artist's impression of the Gaia satellite (image: ESA)

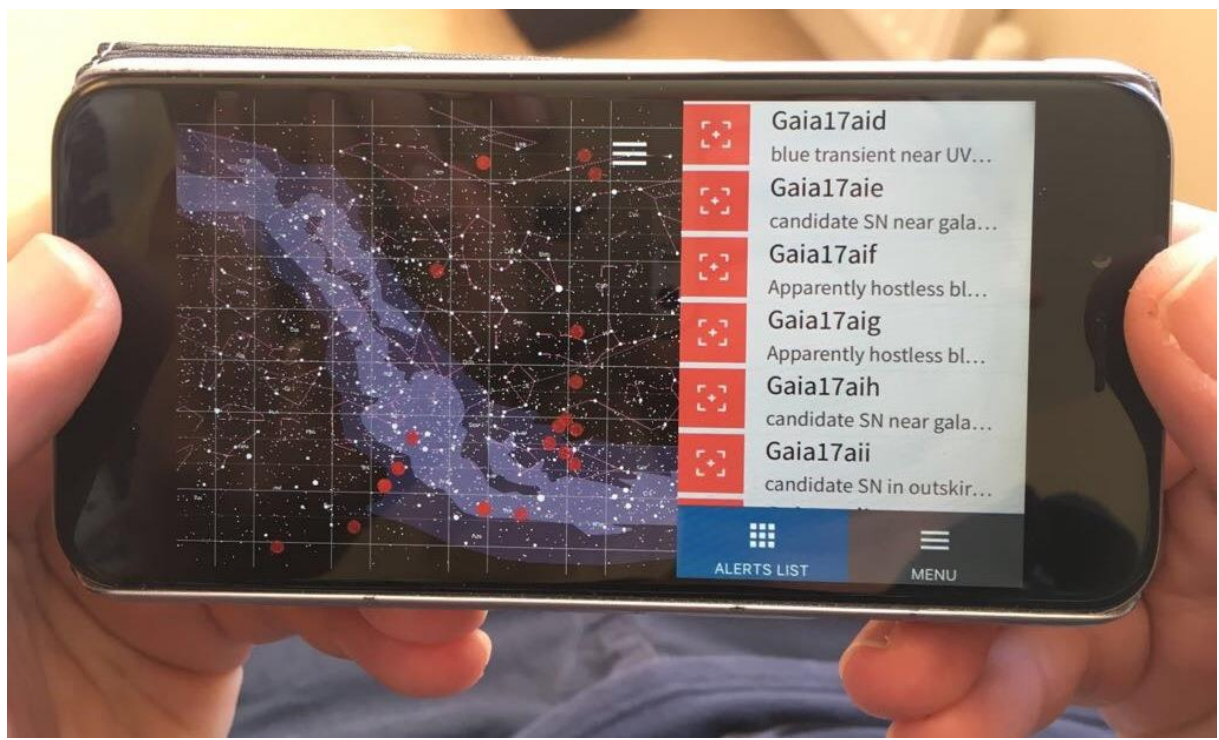

Fig 7 (b): Gaia Alerts via the smartphone App

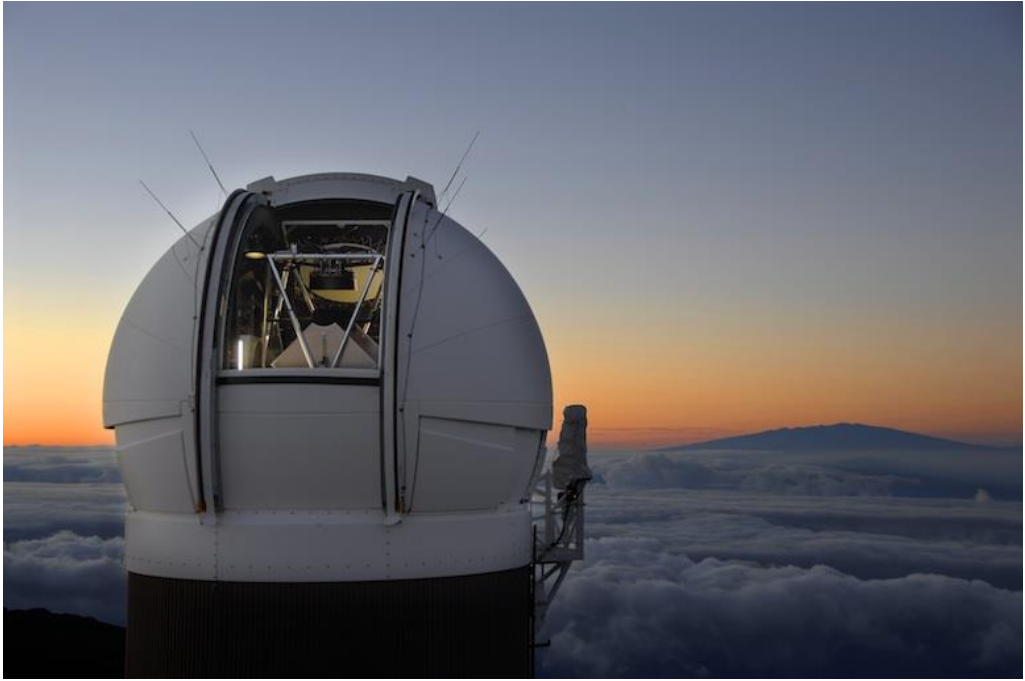

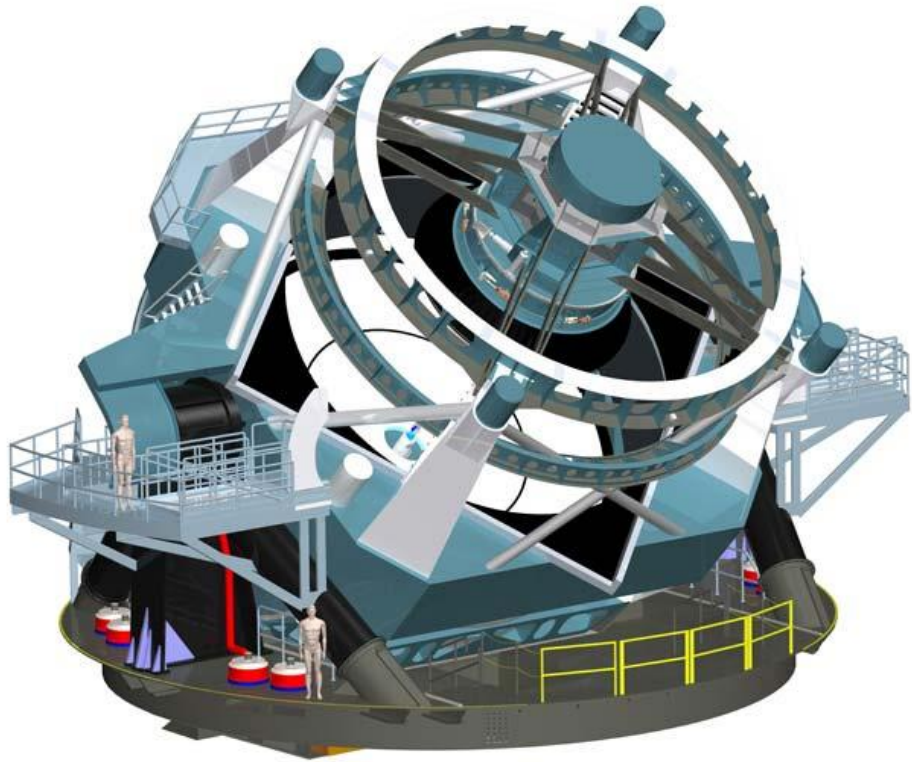

Figure 8: (a) The Panoramic Survey Telescope and Rapid Response System (Pan-STARRS) 1.8 m PS-1 telescope at the summit of Haleakala, Hawaii. (b) Artist's impression of The Large Synoptic Survey Telescope (LSST), a wide-field survey reflecting telescope with an 8.4-meter primary mirror

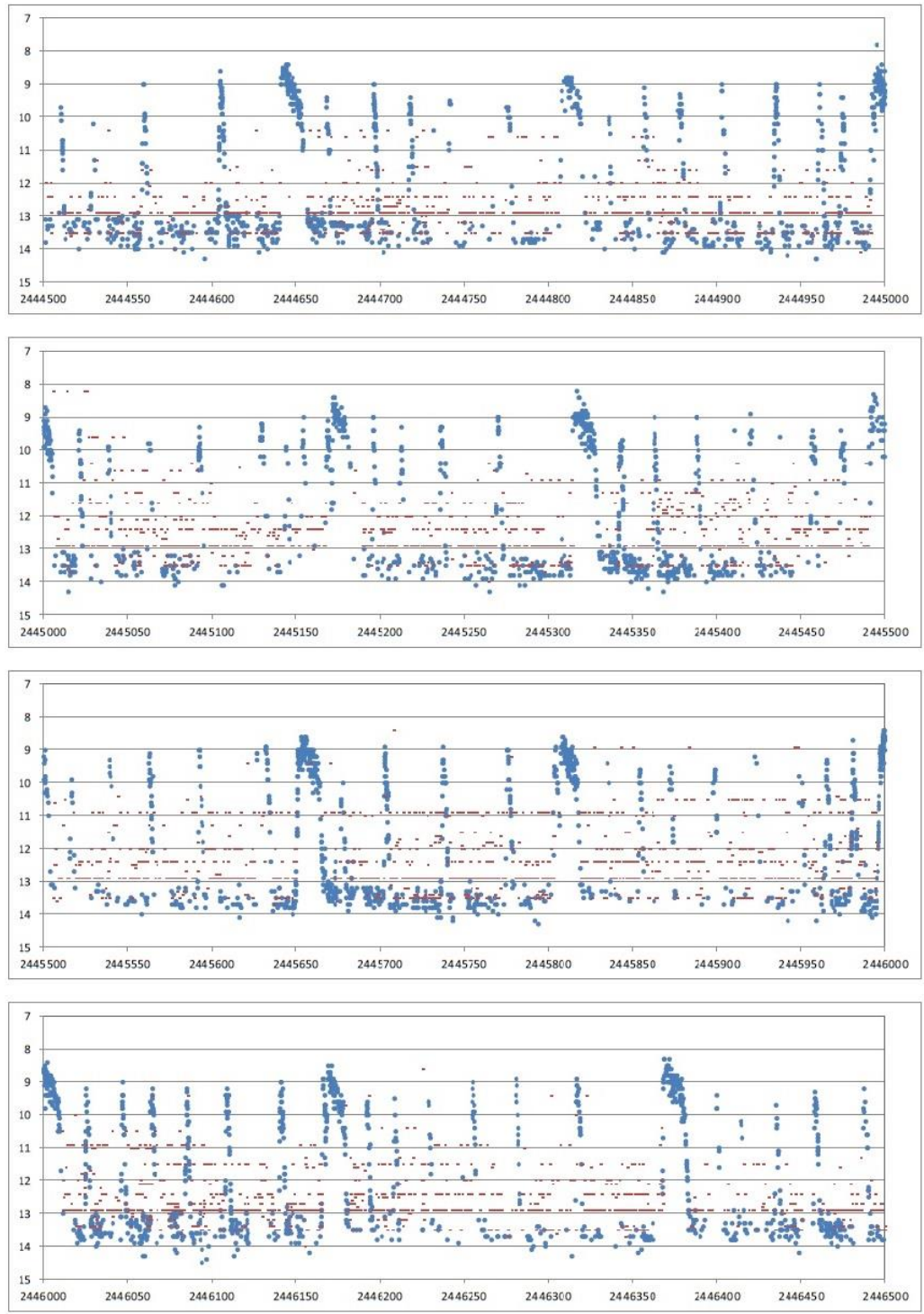

Figure 9: Exquisite coverage of VW Hyi between 1980 Sep and 1986 Mar; Julian date is on the x-axis and magnitude is on the y-axis. Each panel covers 500 days. The longer (and slightly) brighter *superoutbursts* can easily be distinguished from the *normal* outbursts. Red data points are "fainter than" observations when the star was not detected. Data from the AAVSO International Database, largely contributed by RASNZ observers.

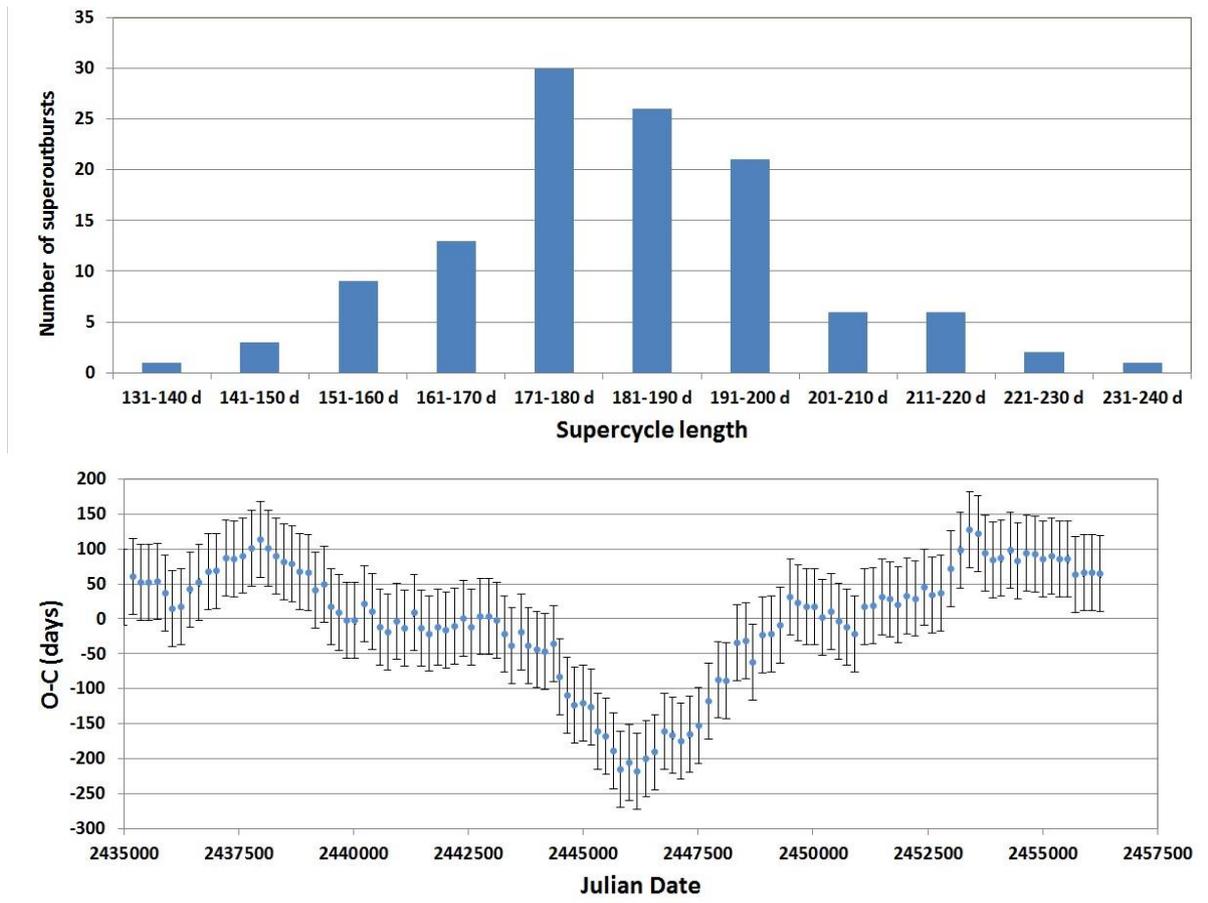

Figure 10. Analysis of the 124 superoutbursts in VW Hyi between 1954 and 2012. (a) Top: distribution of supercycle length. (b) Bottom: Observed minus calculated (O-C) residuals of individual superoutbursts against a linear superoutburst ephemeris of $JD_{max} = 2434765.6955 + 181.58281 \times E$; error bars are the SD of the residuals

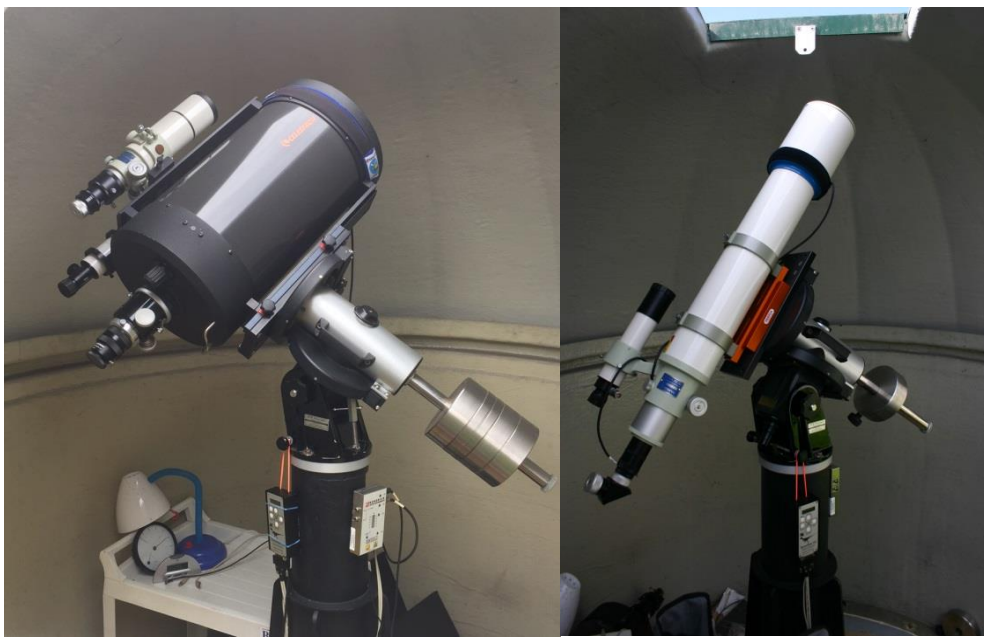

Figure 11: The author's observatory equipped with (a) 28 cm SCT and (b) 10 cm refractor

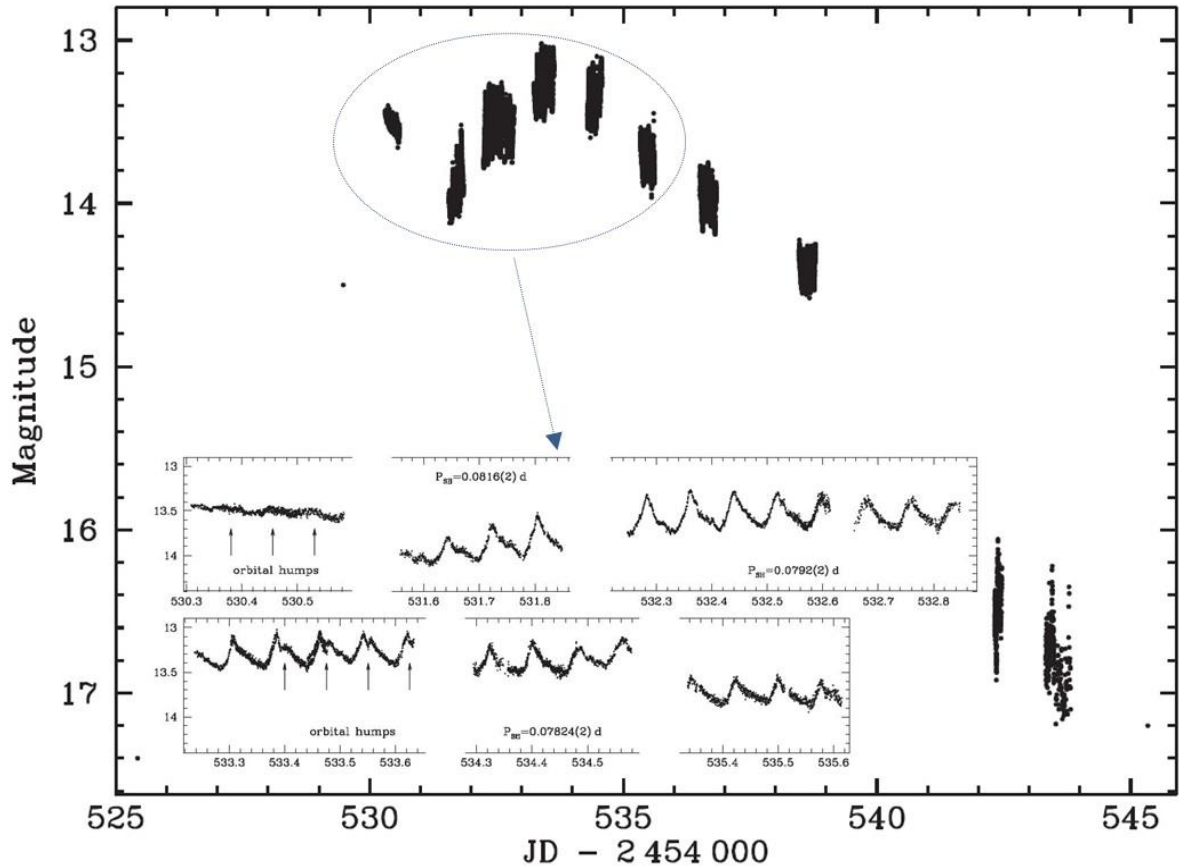

Figure 12: Superoutburst of the SU UMa system, V342 Cam, during 2008 March. The main plot shows the overall profile of the superoutburst (which also shows a precursor outburst at JD529-530 preceding the superoutburst) and inset are details of time-series photometry showing both superhumps, up to 0.4 mag in amplitude, and small orbital humps. Data from Steve Brady, Pavol Dubovský, Ian Miller, Bart Staels and Jeremy Shears. From reference (25)

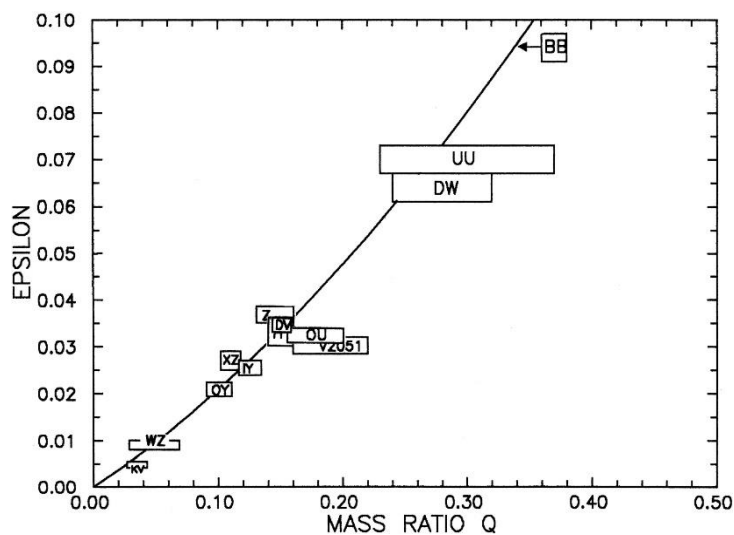

Figure 13: Empirical relationship between the superhump period excess, ε, and q, the mass ratio of secondary star (M$_2$) to white dwarf (M$_1$) resulting in ε = 0.18q + 0.29q$^2$. Boxes are empirical measures of ε and q in several CVs. From Patterson, reference (24)

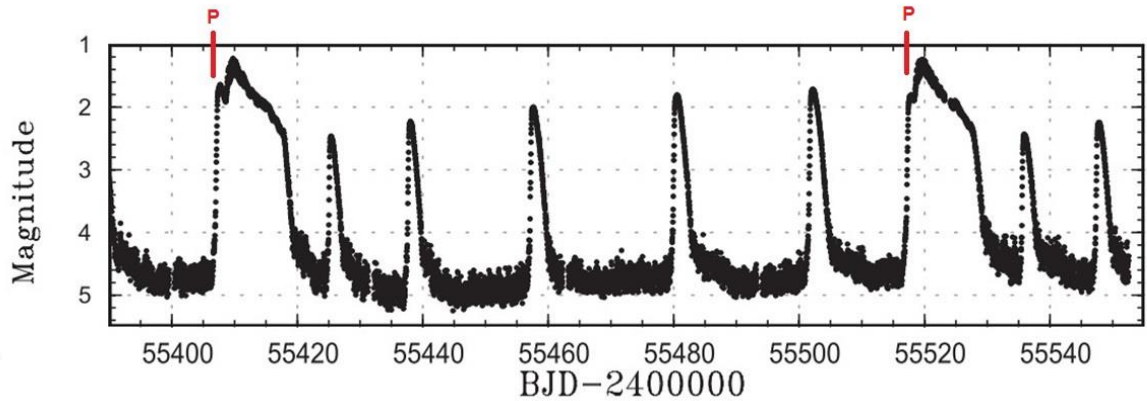

Figure 14: Kepler data for V1504 Cyg showing two superoutbursts, each with a precursor. Note magnitudes are differential magnitudes. From Osaki & Kato (74)

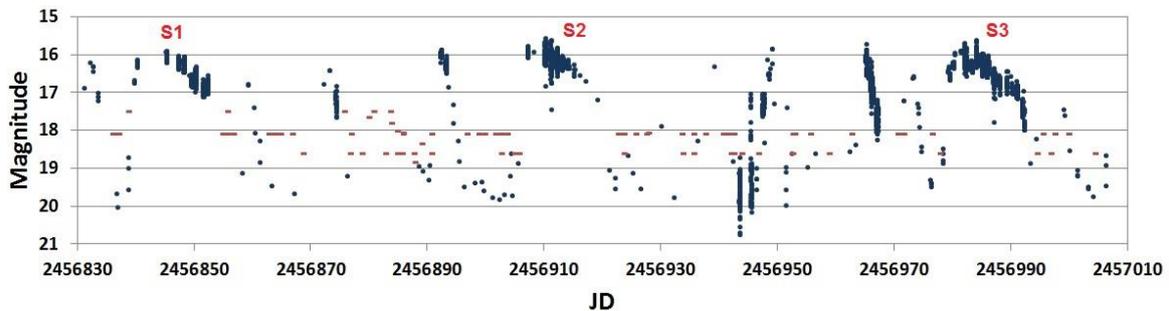

Figure 15: Light curve of CSS 121005:212625+201948 covering 180 days from 2014 June 21 and showing three superoutbursts (S1 to S3). Observers: James Boardman, David Boyd, Denis Buczynski, Pavol Dubovský, Juan-Luis González Carballo, Kenneth Menzies, Ian Miller, Roger Pickard, Gary Poyner, Richard Sabo, Richard Sargent & Jeremy Shears

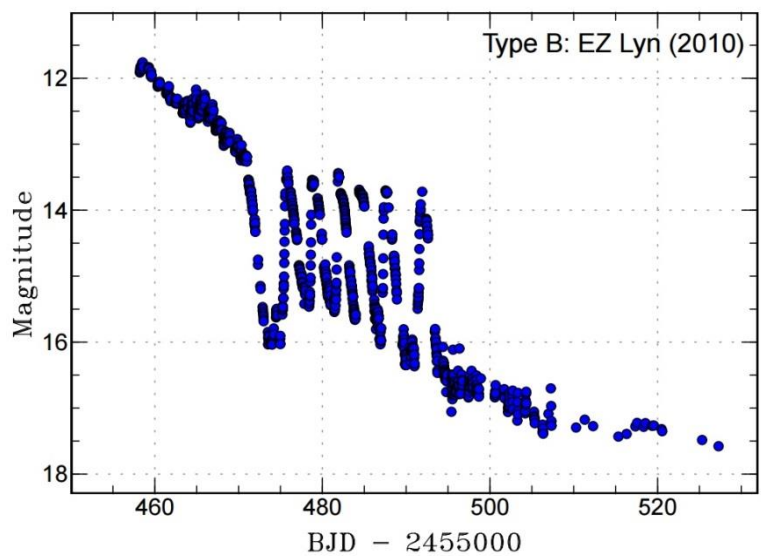

Figure 16: EZ Lyn in 2010 showing six rebrightening episodes following the main outburst. From Kato, reference (28)

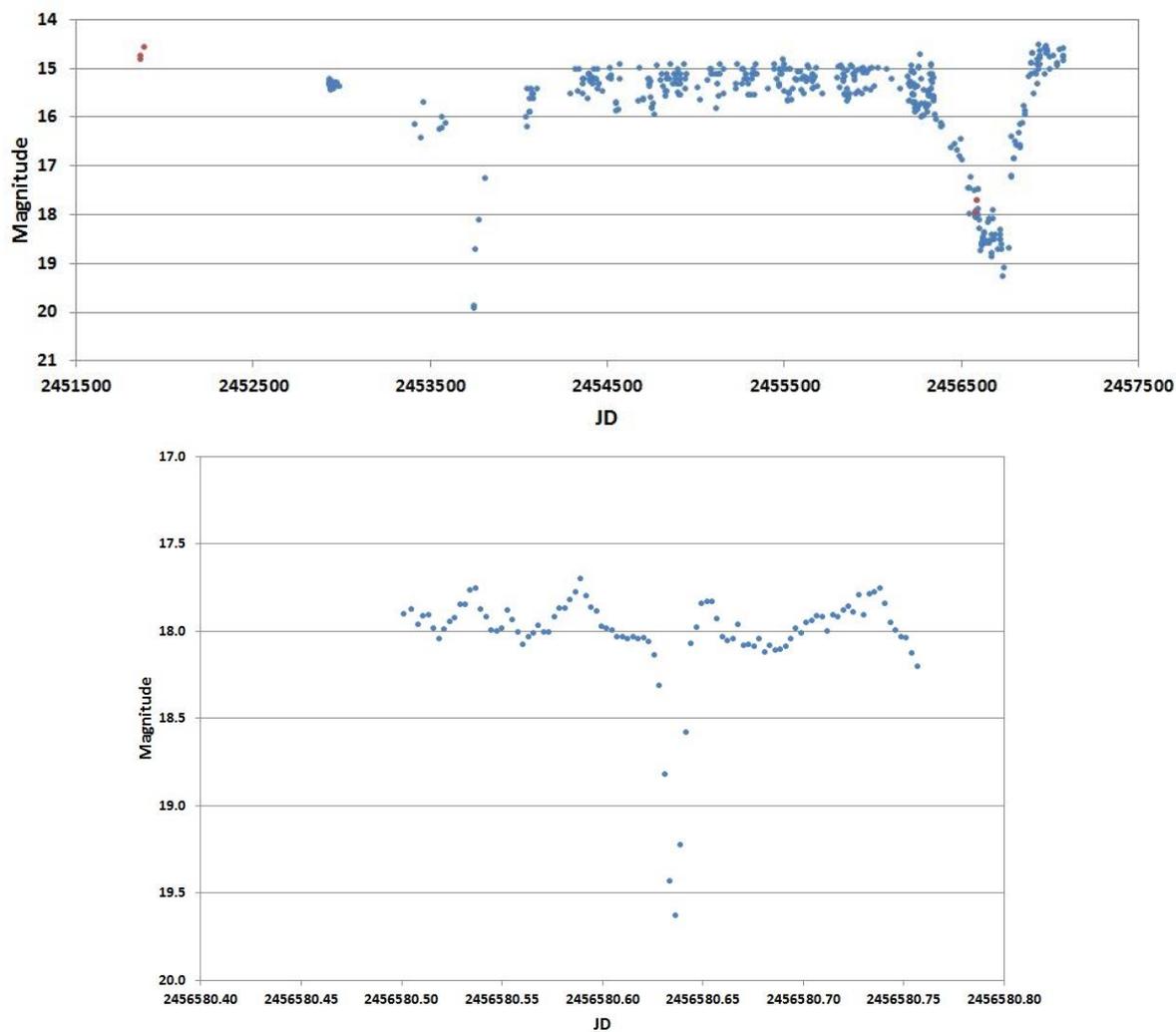

Figure 17: Light curve of HS 0455+8315. (a) Top: from 2000 November to 2015 February. Blue data points are C and V-band data; red data points are R-band. (b) Bottom: Eclipse during the 2013 faint state on October 14 (R-band photometry)

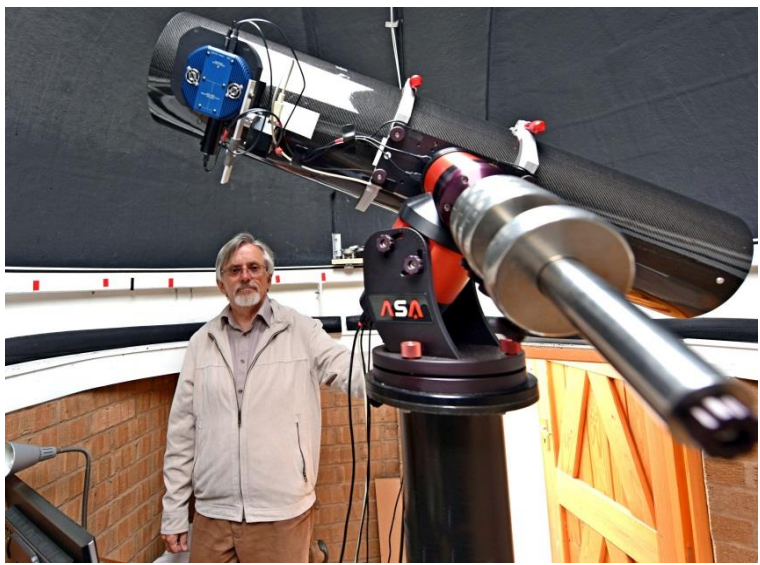

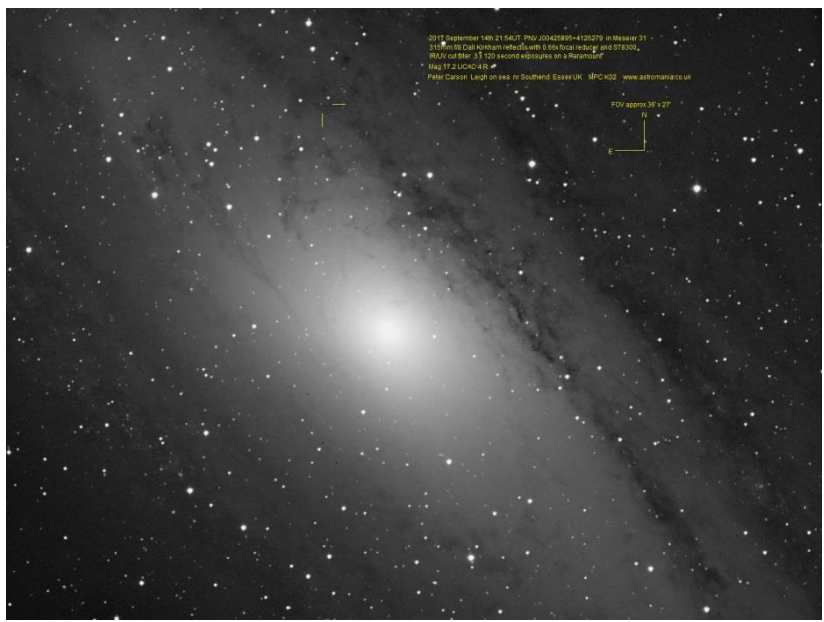

Figure 18: (a) George Carey of Bromsgrove with his 20 cm Newtonian; (b) image of M31 showing George's nova M31N 2017-09b (Peter Carson, Leigh on Sea, 2017 Sept 14, mag 17.2R)

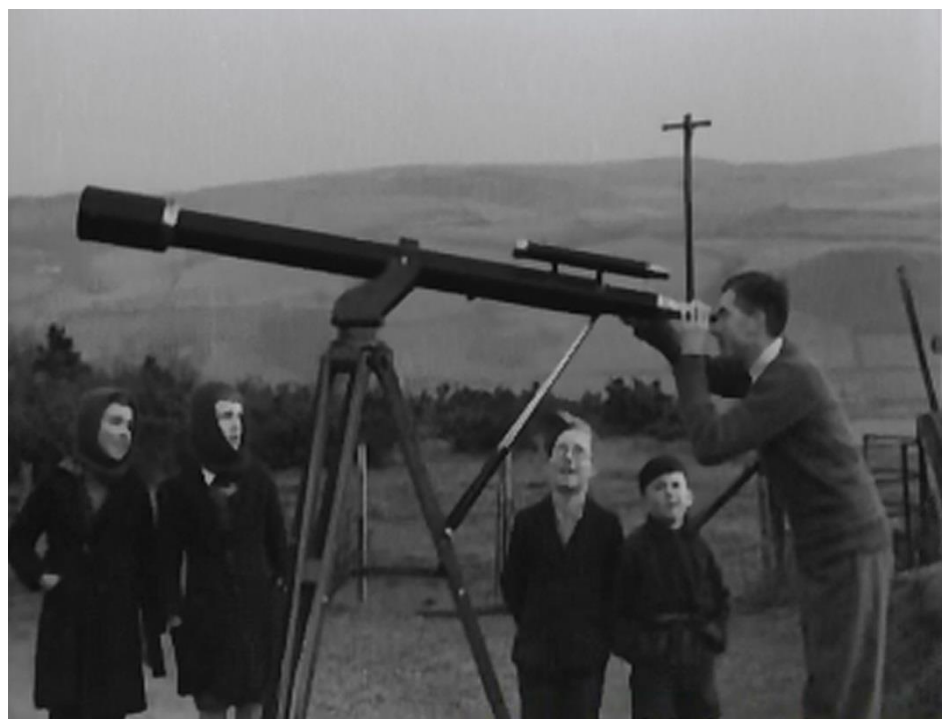

Figure 19: Tony Ellis and his 4-inch (10 cm) refractor

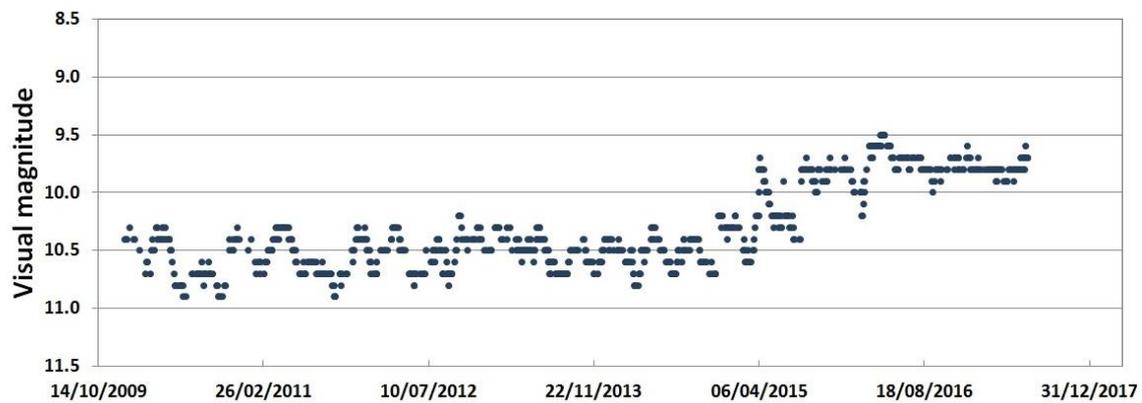

Figure 20: light curve of T Boo showing a super-active state beginning in 2015 Apr. Visual photometry by J. Toone (VSS database)

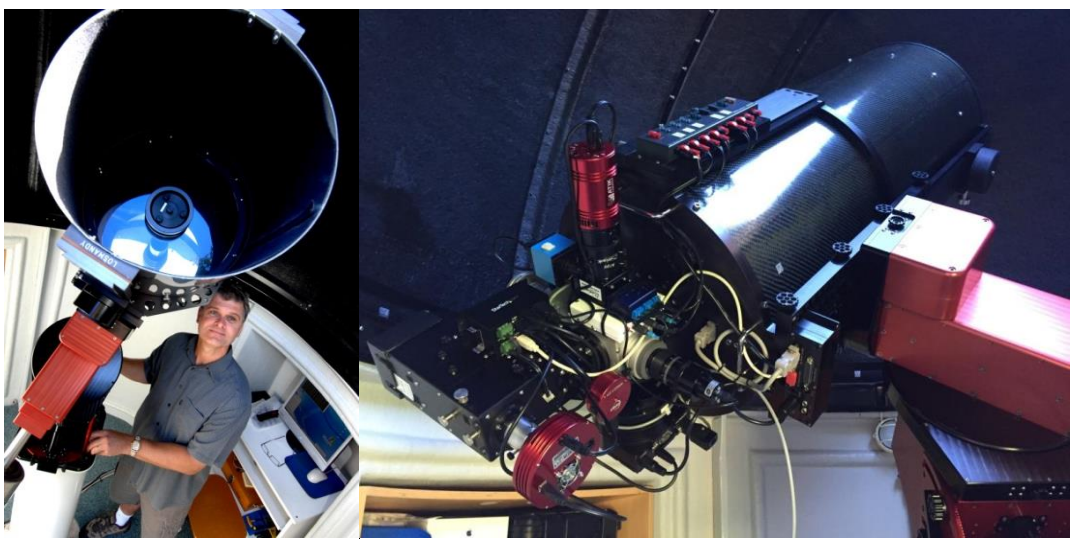

Figure 21: Paul Luckas (left) and his 35 cm telescope complete with spectrograph

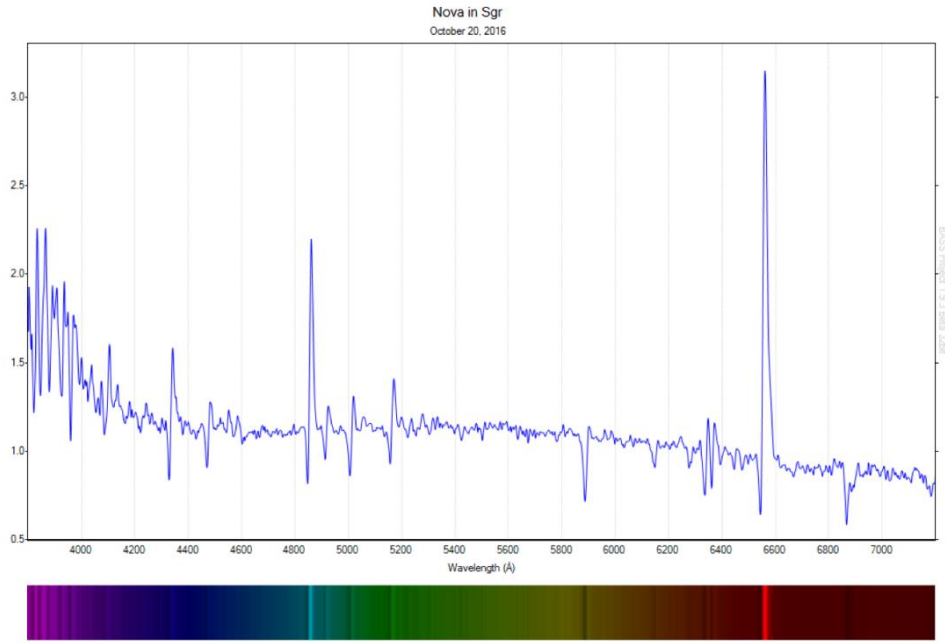

Figure 22: Spectrum of V5855 Sgr (Nova Sgr 2016 number 3) on discovery night, 2016 Oct 20 showing prominent Balmer lines (Paul Luckas)

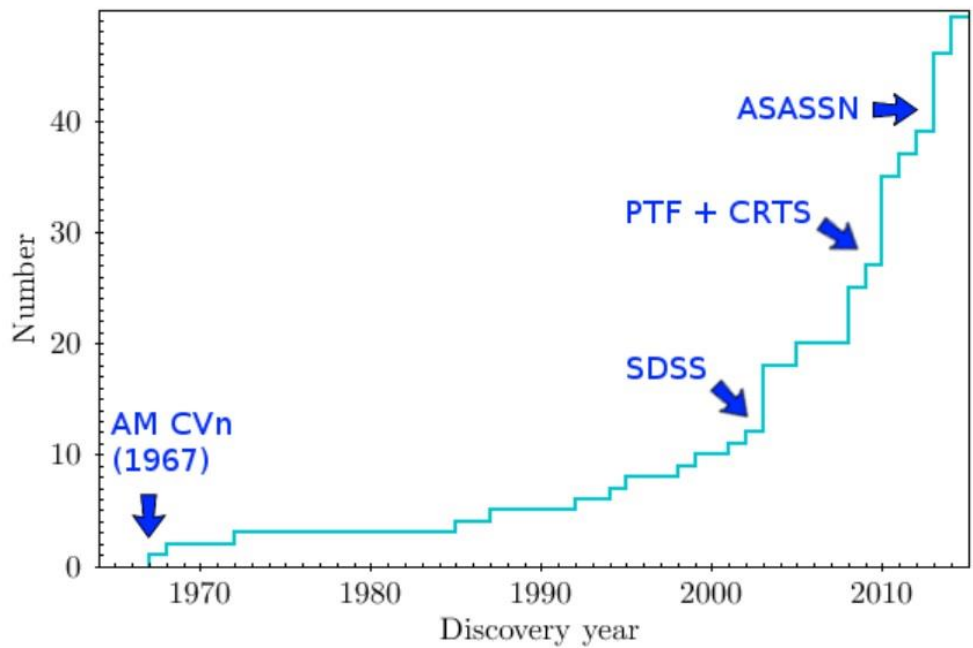

Figure 23: The number of known AM CVn systems until mid-2015, largely driven by survey discoveries: the Sloan Digital Sky Survey (SDSS), the Catalina Real-time Transient Survey (CRTS), the Palomar Transient Factory (PTF), the All-Sky Automated Survey for Supernovae (ASASSN). The start of each of the surveys is marked on the cumulative distribution. From reference (75)

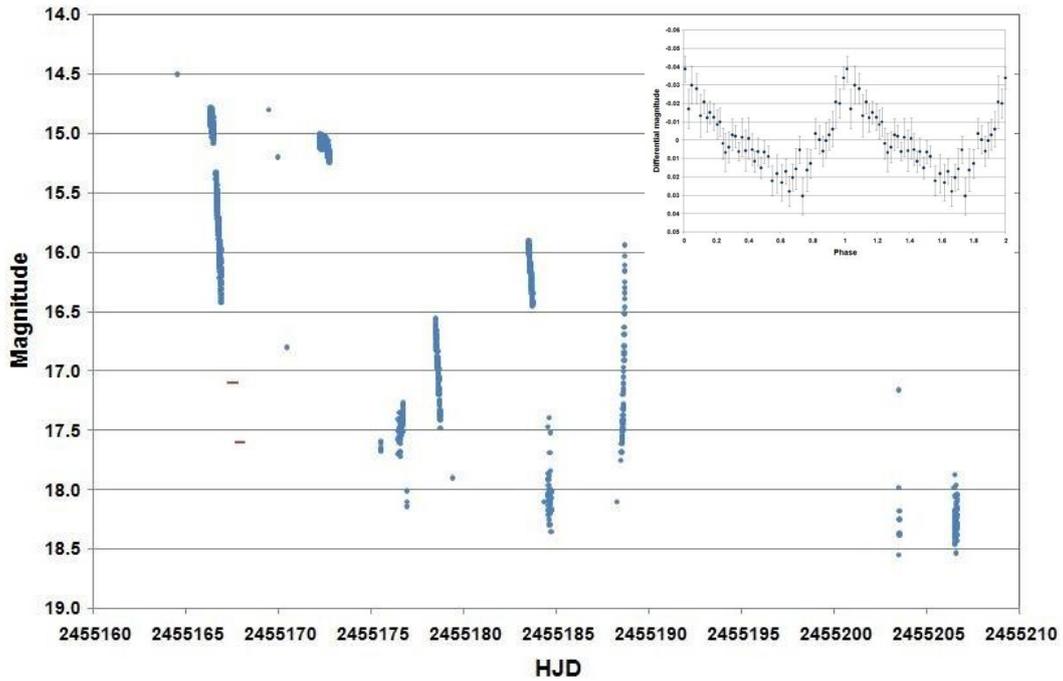

Figure 24: Light curve of the AM CVn systems V744 And during its 2009 outburst showing six rebrightening events. Inset: phase diagram of small (0.06 mag amplitude) but perfectly formed superhumps from one night's photometry

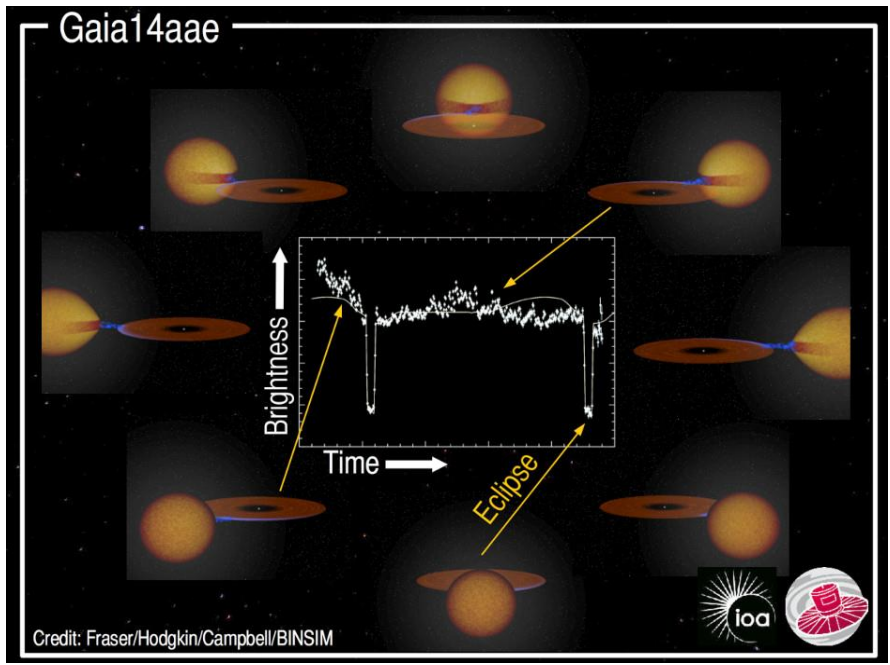

Figure 25: Eclipses in the AM CVn system, Gaia 14aae

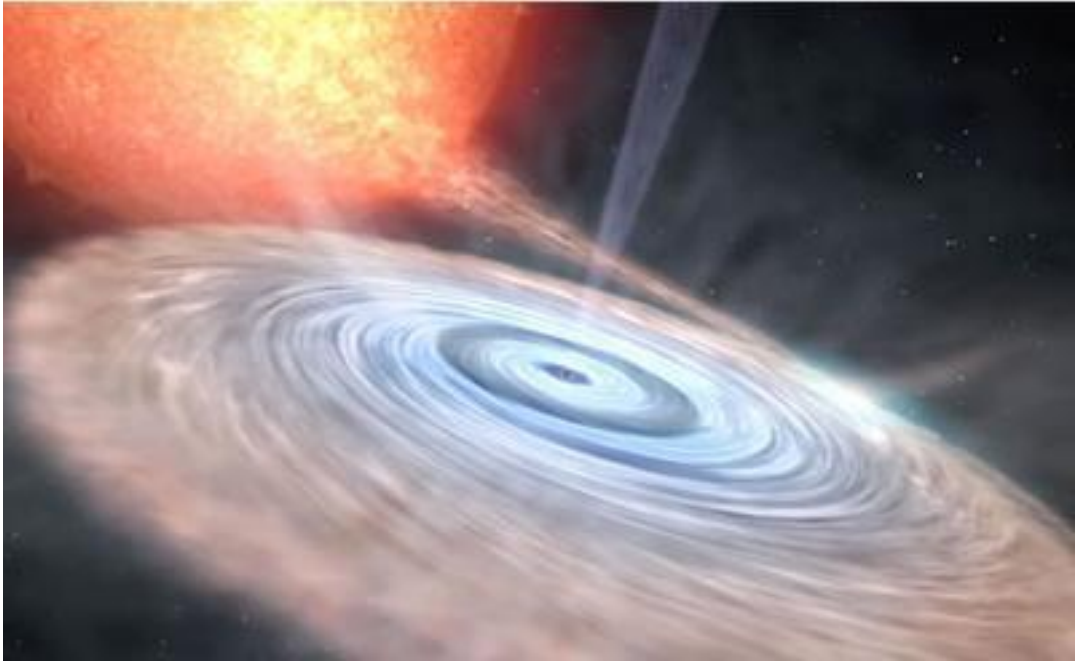

Figure 26: Artist's impression of the accretion disc and perpendicular jets associated with V404 Cygni. Credit: Gabriel Pérez, SMM (IAC)